%% file: main.tex
\documentclass[10pt]{iopart}
\usepackage{lineno}
\usepackage{units}
\usepackage{makecell}
\usepackage{multirow}

\usepackage{scalefnt}
\usepackage[section]{placeins}
\usepackage{lmodern}

\expandafter\let\csname equation*\endcsname\relax
\expandafter\let\csname endequation*\endcsname\relax

\usepackage[version=4]{mhchem}
\usepackage{xcolor}
\usepackage{graphicx}
\usepackage{multicol}
\usepackage{tikz,bm,svg,multirow}
\UseRawInputEncoding 
\usepackage{pgfplots}
\usepackage{hyperref}

\usepgfplotslibrary{groupplots}
\usepgfplotslibrary{dateplot}
\pgfplotsset{ every non boxed x axis/.append style={x axis line style=-},
	every non boxed y axis/.append style={y axis line style=-}}
\usetikzlibrary{arrows.meta}
\usepgfplotslibrary{dateplot} 
\usepgfplotslibrary{groupplots}
\usetikzlibrary{fadings}
\tikzfading[name=fade right, left color=transparent!95, right color=transparent!10, middle color=transparent!95]
\tikzfading[name=fade rect, left color=transparent!95, right color=transparent!70, middle color=transparent!95]
\tikzset{
	semithick/.style={line width=0.8pt},
}

\usepackage{pgfplotstable}

\usepackage{booktabs}
\usepackage{array}
\usepackage{colortbl}
\usepackage{setspace}

\usepackage{pgfplots}
\pgfplotsset{width=7cm,compat=1.15}

\usepackage{subcaption}
\begin{document}

	\title[D retention in Li DIII-D]{Deuterium retention in pre-lithiated samples and Li-D co-deposits in the DIII-D tokamak}
	
	\author{M. Morbey$^{1,2}$, F. Effenberg$^{3}$, S. Abe$^{3}$, T. Abrams$^{4}$, A. Bortolon$^{3}$, R. Hood$^{5}$, U. Losada$^{6}$, A. Nagy$^{3}$, J. Ren$^{7}$, D. L. Rudakov$^{8}$, M. J. Simmonds$^{8}$, D. Truong$^{9}$, and T. W. Morgan$^{1,2}$}
	\address{$^{1}$ DIFFER-Dutch Institute for Fundamental Energy Research, De Zaale 20, 5612 AJ Eindhoven, The Netherlands}
	\address{$^{2}$ Eindhoven University of Technology, De Zaale, Eindhoven, 5612AZ, The Netherlands}
 \address{$^{3}$ Princeton Plasma Physics Laboratory, Princeton, NJ 08543, USA}
 \address{$^{4}$ General Atomics, San Diego, CA 92186, USA}
 \address{$^{5}$ Sandia National Laboratories, Livermore, CA 94551, USA}
 \address{$^{6}$ Auburn University, Auburn, AL 36849, USA}
 \address{$^{7}$ UTK - University Tennessee, Knoxville, TN 37996, USA}
 \address{$^{8}$University of California - San Diego, La Jolla, CA 92093, USA}     
 \address{$^{9}$Lawrence Livermore National Laboratory, Livermore, CA 94550, USA}
	
	\ead{m.morbey@differ.nl}
	\vspace{20pt}
	\begin{indented}
		\item[]June 2024
	\end{indented}
	
	\begin{abstract}

Divertor designs involving liquid lithium have been proposed as an alternative to solid designs and wall conditioning techniques. However, Li affinity with tritium poses a risk for the fuel cycle. This study investigates deuterium retention in pre-lithiated samples and Li-D co-deposits in the DIII-D tokamak, making for the first time a direct comparison between Li-D co-deposits and pre-deposited Li films. Samples were exposed to H-mode plasmas in the far scrape-off layer (SOL), and Li powder was injected in-situ with the impurity powder dropper to study the uniformity of Li coatings, and the dependence of fuel retention on Li thickness. The results show that at temperatures below the melting point of lithium, deuterium retention is independent of the thickness of pre-deposited Li layers, with Li-D co-deposits being the primary factor for fuel retention. Both pre-deposited and in-situ deposited Li showed lower erosion than predicted by sputtering yield calculations. These results suggest that fuel retention in fusion reactors using lithium in the divertor will likely be dominated by co-deposits rather than in the divertor itself. If one desires to use Li to achieve flatter temperature profiles, operando Li injection is advantageous over pre-deposited Li films, at least at temperatures below the melting point of lithium. 

\end{abstract}

	\ioptwocol
	
	\section{Introduction}
	
Plasma interactions with plasma-facing components (PFCs) represent a critical challenge for future fusion reactors \cite{you2022divertor,romanelli2012fusion}. On the material side, the divertor, subjected to high heat and particle flux, is prone to recrystallization, erosion, or melting. Conversely, from the plasma perspective, PFCs can act as an undesired source of impurities through chemical and physical sputtering, posing a significant risk for radiative collapse, especially in fully high-Z wall tokamak designs like SPARC and ITER \cite{rodriguez2022overview, loarte2024initial,zamperini2023separatrix}.

To address the high heat fluxes anticipated in the divertor of future reactors, liquid metals have been proposed as an advance material solution, with a wide range of designs outlined in overview papers \cite{nygren2016liquid, tabares2015present}. Liquid metals offer several advantages, including self-replenishment and the ability to operate in a regime where incoming power is decoupled from the target temperature as it is shown and explained in  \cite{morgan2017liquid}. One such proposed liquid metal is lithium (Li), valued for its low melting point and low atomic number \cite{ono2017liquid}. Furthermore, Li powder injection has been used in several tokamaks for wall conditioning, Edge Localized Mode (ELM) suppression, high heat flux handling and access to low recycling regimes \cite{maingi2011continuous,sun2019real,osborne2015enhanced,effenberg2022mitigation, boyle2023extending}.

Despite the several advantages of Li, fuel retention in Li walls raises concerns. Li retains hydrogenic species in ratios up to 1:1 Li:H \cite{baldwin2002deuterium, veleckis1974lithium, morbey2024deuterium}. Given the limited inventory of tritium (T) and its radioactive nature, it is crucial to minimize the amount of tritium  trapped in the vessel walls, in  order to decrease the demand on the tritium fuel cycle. T retention is especially concerning in regions that are difficult to access.  

When Baldwin et al. \cite{baldwin2002deuterium} exposed to a deuterium (D) plasma, they observed a 1:1 Li:D ratio  in liquid Li. In the observations in Ou et al. \cite{ou2022deuterium}, it was concluded that this ratio is only observed in the top Li layers of pre-filled Li targets.
It is believed that as Li-D forms on the surface, it precipitates, potentially creating a barrier that prevents D from diffusing deeper and bonding with the bulk Li. In the co-deposition regime, with Li and D depositing on the wall simultaneously, there is no diffusion barrier, in contrast to the pre-lithiated cases. While fuel retention studies have been conducted on Li targets and Li-H co-deposits, there is a lack of direct comparison between the two regimes and the dependence of hydrogen retention on the thickness of the pre-deposited Li layer. This distinction is particularly important for assessing the most suitable Li technique for optimal wall conditioning and access to low recycling regimes.

In this study, deuterium retention in pre-lithiated samples of varying thicknesses and in Li-D co-deposits is compared by exposing the samples to H-mode D plasmas in the DIII-D tokamak with in-situ lithium powder injection.
	
	\section{Experimental setup for lithium injection and divertor collection}\label{sec:exp_method}
\subsection{Sample fabrication, characterization techniques and pre-plasma characterization results}
 Three types of samples were used in this work: bulk stainless steel (SS) samples, graphite samples with tungsten (W) coatings (C+W) and bulk SS samples with Li coatings (SS+Li). The surface composition of the prepared samples prior to plasma exposure was determined at Dutch Institute for Fundamental Energy Research (DIFFER) using Ion Beam Analysis (IBA) on reference samples that were not exposed to plasma. Specifically, Elastic Backscattering Spectroscopy (EBS) and Nuclear Reaction Analysis (NRA) were employed to measure the carbon (C) and oxygen (O) content, as well as the thickness of the deposited lithium and tungsten layers. For EBS, a 2.1 MeV proton beam was used, while NRA was conducted with a 2.3 MeV $^3$He beam. For both techniques, the ion beam was incident at an angle of 0$^\circ$ relative to the surface normal of the sample, with a beam spot size of 1.5 mm $\times$ 1.5 mm and the scattering angle was 170$^{\circ}$. For EBS analysis, the lithium areal density was quantified using the cross-section measured in \cite{paneta2012differential}, while the oxygen and carbon cross-sections were calculated using SigmaCalc \cite{gurbich2016sigmacalc}. The cross-sections used to analyze the D and Li peaks from NRA were taken from \cite{wielunska2016cross} and \cite{bondouk1975experimental}, respectively. The ion beam spectra were analyzed using the DataFurnace (NDF) software \cite{barradas1997simulated}.
 
 Two sets of seven samples were used in these experiments. Four of the installed samples were made of bulk polished Stainless Steel (AISI 304 / 1.4301). The roughness of the non-exposed samples was assessed using a profilometer (contourGT-I from Bruker). These bare SS samples showed a root-mean-square roughness of $\mathrm{R_{q}}$=0.07 $\mu$m over a 0.24 mm $\times$ 0.18 mm area.
 As a case study for C machines transitioning to W PFCs by W deposition on C tiles, five graphite samples with 1 $\mu$m or 1.6 $\mu$m W coatings were used. This type of sample did not undergo such an extensive polishing method as the SS samples, as rougher surfaces have been proven to increase layer adhesion \cite{matvejivcek2013influence,van2020effect}. The W coatings were manufactured via magnetron sputtering at General Atomics (GA).  The C + 1 $\mu$m W sample revealed an $\mathrm{R_{q}}$ of 0.530 $\mu$m over a 0.24 mm $\times$0.18 mm area. Roughness data for the C + 1.6 $\mu$m W sample is unavailable. In contrast, an $\mathrm{R_{q}}$ of 0.690 $\mu$m was recorded for a backup graphite sample with a 0.3 $\mu$m W coating, suggesting that the W coating mitigates the roughness of the graphite substrate.
 The thickness of the W coating on the carbon substrates prior to plasma exposure was confirmed by EBS. The non-exposed samples exhibited a W thicknesses of 1.04 $\mu$m and 1.66 $\mu$m for the C + 1 $\mu$m W and for the C + 1.6 $\mu$m W, respectively. However, EBS revealed that the W layer already contained 10$\%$ C and 10$\%$ O, attributed to the inherent roughness of the substrate and the affinity of W to react with O. Energy Dispersive X-Ray (EDX) analysis of these same non-exposed C+W samples showed that the W coating had a high carbon and oxygen content, around 25$\%$ and 13$\%$, respectively, which is consistent with the EBS results showing that the deposited layer is not a pure W layer. Given the higher accuracy of EBS over that of EDX with respect to lower atomic number elements, we use the EBS results as the reference for the pre-plasma impurity content of the C+W samples.

\begin{table}[h!]
    \centering
    \begin{tabular}{c|c|c|c|c}
     \multirow{2}{*}{ Sample}&  EBS& Profl. & \multicolumn{2}{c}{EBS/EDX} \\ 
    & {W [\(\mu m\)]} & { $\mathrm{R_{q}}$ [\(\mu m\)]} & {O [\%]} &{C [\%]}  \\ 
    \hline
     C + 0.3\(\mu m\)W & - & 0.690 & -&-\\  \hline
    C + 1\(\mu m\)W & 1.04 & 0.530 & 12/13 & 10/25 \\ 
    \hline
    C + 1.6\(\mu m\)W & 1.66 & - & 12.5/- & 12/ - \\ 
    \end{tabular} 
    \caption{Summary of the C+W reference samples. Deposited W thickness measured by EBS, root-mean-square roughness from a profilometer, and the C and O composition percentages in the deposited W layer measured by EBS and EDX.}
    \label{tab:coating_properties}
\end{table}

 To study the difference between D retention in LiD co-deposits and pre-deposited Li coatings, six of the stainless steel samples were pre-coated with Li at Princeton Plasma Physics Laboratory (PPPL) by evaporation, under a base pressure of 10$^{-9}$ Torr, where the thickness of Li was monitored using a quartz crystal microbalance (QCM). Although the lithium-coated SS samples were stored in argon (Ar) filled plastic bags during transport, air exposure occurred when the samples were unmounted from the Li deposition chamber, leading to surface oxidation prior to plasma exposure. Three samples of each thickness, 350 and 750 nm, were fabricated, and one of each of them was used to characterize the pre-plasma exposure Li thickness using EBS. Having pre-lithiated samples with different thicknesses, allowed us to study the influence of the pre-deposited Li thickness on D retention.

    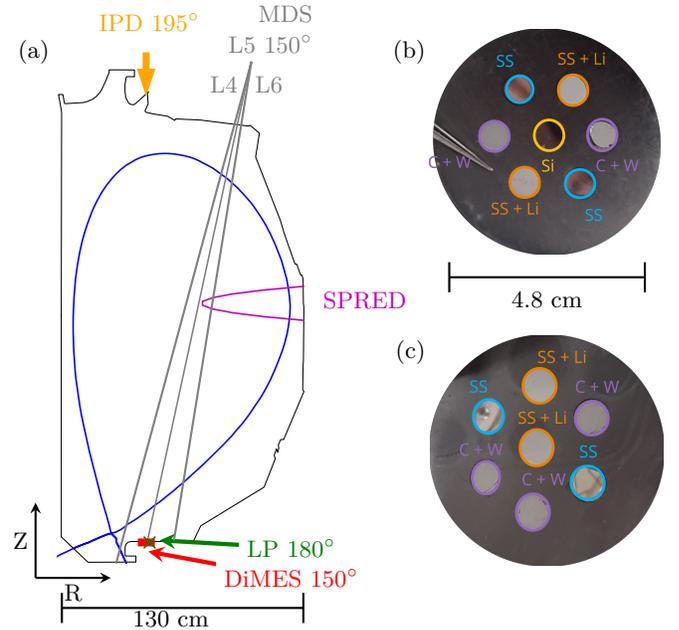
\begin{figure}
        \centering      
        \input{Pre-plasma-figures}
        \caption{(a) DIII-D poloidal cross-section with magnetic
equilibrium separatrix using EFIT01 of shot $\#$196020 at t=3.5 s. In this figure, the main diagnostics are represented including their
toroidal angle. (b) and (c) DiMES head 1 and 2 before plasma exposure, respectively.} 
        \label{fig:DiMES-heads}
    \end{figure}

    After plasma exposure samples were handled and transferred in Ar-filled containers and gloveboxes. Composition analysis after plasma exposure was performed at DIFFER via NRA and EBS, in a similar way as described for the pre-exposure surface composition analysis, or at the University of California San Diego (UCSD) via thermal desorption spectroscopy (TDS). TDS was performed up to 900 $^\circ$C with an average heating rate of 4.84 $^\circ$C/s.

 \subsection{DIII-D setup and diagnostics for lithium powder injection experiments}

      
 Figure \ref{fig:DiMES-heads} (a) shows the separatrix magnetic equilibrium reconstruction, using EFIT01, a code that performs the equilibrium Grad-Shafranov equation using solely measurements from magnetic diagnostics \cite{lao1985reconstruction}.  The impurity powder dropper (IPD) which uses piezoelectric actuators to vibrate a pathway, delivering a controlled powder flow from a reservoir into the plasma was used to drop Li powder with a diameter of 45 $\mu$m during plasma \cite{nagy2018multi, MANSFIELD2010890}. The IPD is located at the top of the DIII-D tokamak (major radius R = 1.485 m and $\phi$ = 195$^{\circ}$ ). It is shown by modelling that the injected particles are ablated close to the separatrix \cite{bortolon2020observations, EFFENBERG2025101832}. The Divertor Material Evaluation System (DiMES)  in the lower divertor at R=1.47-1.52 m at 150$^{\circ}$ was used to expose samples in DIII-D \cite{wong1992divertor}. The main experiment was conducted in two parts: the first focused on ELM suppression by lithium with the first sample holder (head 1) mounted, and the second examined the effect of lithium in Resonant Magnetic Perturbations (RMP) scenarios using the second sample holder (head 2). 
    In Figures \ref{fig:DiMES-heads} (b) and (c) the 6 mm diameter button samples mounted on the first and second DiMES heads are shown. The sets differ only in one sample, the first set of samples included a Si sample with a micro-trench, which is not analyzed in this work, while the second set of samples consisted of the same samples as the first but instead of the Si sample, a third C + 1 $\mu$m W sample was installed. An overview of the samples is shown in Table \ref{tab:Li_thickness}. The shots for these experiments were $\#$196018 to $\#$196033.
    Head 1 was inserted during shots $\#$196018 to $\#$196026. These samples were exposed to 5 shots with Li injection ($\#$196022-$\#$196026) corresponding to a total of 111 mg of Li. Samples in head 2 were mounted between shots $\#$196027 and $\#$196033 and exposed to 3 shots with Li injection ($\#$196030, $\#$196031, $\#$196033), corresponding to a total of 142 mg of injected Li. 
    During plasma, part of the injected Li was deposited on the samples mounted in DiMES. Given the chosen magnetic configuration, the samples were exposed to the far scrape-of-layer (SOL) plasma.
    
    During plasma operation, a Multi-chord Divertor Spectrometer (MDS) was used to monitor Li-$\mathrm{{II}}$ line intensity ($\lambda$= 548 nm) at DiMES (view L5) and across the lower divertor (L views), enabling to determine an estimate of the Li$^{1+}$ flux, $\Gamma_\mathrm{{Li^{1+}}}$ flux at the sample locations using the S/XB method, which is discussed in detail in section \ref{sec:sxb}. Surface Langmuir probes (LP) located at a toroidal angle $\phi$= 180$^{\circ}$ both on the shelf at R=1.51 m and R=1.61 m as well as on the floor at R=1.31 m were used to determine the plasma conditions at DiMES/ MDS L5 view, MDS L6 view and MDS L4 view, respectively \cite{watkins2008high,brooks1992multichord}. Li-$\mathrm{{III}}$ at $\lambda=11.4$ nm emission from the plasma core was monitored with the vacuum ultraviolet spectrometer SPRED \cite{fonck1982multichannel,mclean2018quantification}.

 \section{Effects of lithium injection in H-mode plasmas}
 For these experiments, lower single null (LSN) H-mode plasmas with an ITER-like shape were achieved using a neutral beam power of $\mathrm{P_{NBI}= 3.5 }$ MW and on-axis magnetic field at $\mathrm{B_{t}=2}$ T. The plasma current was $\mathrm{I_P}=1.2$ MA. The flat top energy stored in the plasma was 0.6 MJ and line-averaged density in the core was $\mathrm{n_{e}= 4.5\times10^{18}}$ m$^{-3}$. The Li injection rate was varied between 2.6 mg/s and 20.6 mg/s for the ELM suppression focus of the experiment.  The time evolution of several measurements is shown in Figure \ref{fig:plasma_param} for three shots, shot $\#$196020, a reference shot before any Li injection and with DiMES retracted, shot $\#$196021 before any Li injection and with DiMES inserted, and shot $\#$196024, a shot with Li injection. This figure illustrates Li signals from SPRED, Balmer-alpha emission at $\lambda=$ 656.27 nm (D${\mathrm{\alpha}}$) and flux of deuterium ions $\Gamma_{\mathrm{D^{+}}}$ from LP and lithium ion flux $\Gamma_{\mathrm{Li^{+}}}$ at DiMES measured by L5 view of MDS, with a 2.54 cm diameter viewing size centered in the middle of the sample holder, which if perfectly aligned would see part of all samples. Li injection started at 1.2 s but only reaches the plasma core at 2.2 s, as seen with mid-plane SPRED in Figure \ref{fig:plasma_param} (a). In Figure \ref{fig:plasma_param} (b) shows a $25\%$ decrease in the mean amplitude of D$_{\alpha}$ signal following Li injection, indicating a reduction in recycling. Langmuir probe measurements show no effect of Li injection on $\Gamma_{\mathrm{D^{+}}}$, Figure \ref{fig:plasma_param} (c). Finally, in Figure \ref{fig:plasma_param} (d) the L5 view from MDS exhibits the $\Gamma_\mathrm{Li{+}}$ proximate to the sample locations. Notably, an increase in the Li signal here correlates with an increase in the Li emission detected by SPRED, suggesting a relationship to the deposition of Li powder. This phenomenon will be discussed in the subsequent section.

    \section{Quantifying lithium erosion using the S/XB method} \label{sec:sxb}
\begin{figure}
     \centering
    \input{plasma_parameters}
     \caption{ Time evolution of H-mode discharges for reference shot, $\#$196020, reference shot with DiMES retracted, $\#$196021, reference shot with DiMES inserted and a shot with Li injection, $\#$196024. The purple area indicates the time span used for S/XB analysis, after flat top is reached and before Li enters the plasma. (a) Time evolution of Li$\mathrm{_{III}}$  emission measured with SPRED. (b) Time evolution of the D$_{\alpha}$ signal, a reduction is observed when Li is injected indicating reduced recycling. D$_{\alpha}$ was measurements were no available for shot $\#$196021 (c) Time evolution of D flux measured by the Langmuir probes close to DiMES, showing that D flux was similar without and with Li injection, (d) Li flux determined by the S/XB method using data from view L5 of MDS in logaritimic scale, where it is observed that the Li signal from pre-lithiated samples is negligible, it is similar with and without DiMEs inserted.}
     \label{fig:plasma_param}
 \end{figure}
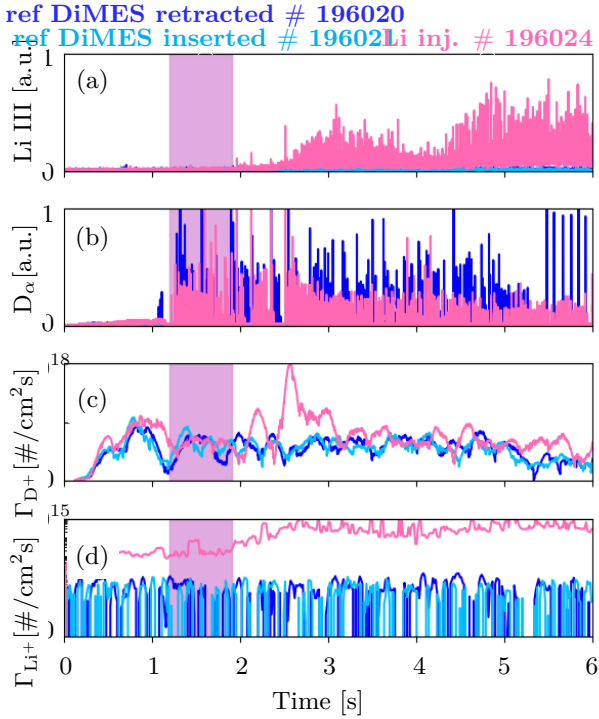

    Prior to any Li powder injection, the L5 view of MDS revealed no detectable Li signal, as seen in Figure \ref{fig:plasma_param} (d). This absence allows us to confidently attribute any subsequent observed Li signal in the MDS data to the injection of Li powder and not from the pre-deposited Li in the samples. After the injection of 111 mg of Li powder, a new set of samples (head 2) was exposed to reference shots without additional Li powder injection. Analysis showed a lower lithium emission compared to during injection, but with a higher baseline level than before any Li injection. Similarly elevated baseline Li levels were observed in views near the DiMES location, specifically in L4 and L6. This consistency in the Li signal across multiple views indicates that the signal is due to legacy lithium from previous injections, providing further assurance that the Li originating from the pre-lithiated samples does not experience significant sputtering. 
    To quantitatively determine the in-situ Li erosion from Li deposits (from the powder dropper) we look at the time span between reaching flat top and beginning of Li injection for each shot, between 1.2 s and 1.9 s. The S/XB method was used \cite{pospieszczyk2010determination}.  Flux densities were calculated using equation \ref{eq:1} \cite{abrams2021evaluation} and plotted in Figure \ref{fig:plasma_param} (d):
    \begin{equation}
\Gamma_{\mathrm{Li}}=\frac{4 \pi}{1-r_{\mathrm{Li}}}\left(\mathrm{\frac{S}{X B}}\right)_{\mathrm{Li}_\mathrm{II}} I_{\mathrm{Li}_\mathrm{II}}
\label{eq:1}
\end{equation} 

where $\mathrm{\left(\frac{S}{X B}\right)_{\mathrm{Li}_\mathrm{II}}}$ is the ionization/photon coefficient from the ADAS database \cite{summers2007adas}, $I_{\mathrm{Li}_\mathrm{II}}$ is the measured spectral $\mathrm{Li}_\mathrm{II}$ line emission intensity and $r_{Li}$ is the fraction of $\mathrm{Li}^{+1}$ promptly-redeposited before following ionization. Given that the samples were exposed to electron densities $\mathrm{n_{e}, \text { div }}=1-4 \times 10^{18} \mathrm{~m}^{-3}$, electron temperatures of $\mathrm{T_{e, \text { div }}=20-30 }$ eV, and toroidal magnetic field $ \mathrm{B_t = 2~T}$, this fraction is estimated to be $\mathrm{r_{Li}\simeq0.02}$ \cite{naujoks1996tungsten}.

The erosion of the in-situ Li coating was assessed for the shots following the first shot of Li deposition on each head ($\#$196023-196029 $\#$196031-196033), the erosion rate was found to be between $3\times 10^{13} $ - $5\times 10^{14}  \mathrm{\#/cm^{2}s}$. Integrating the erosion rate by the time interval corresponding to the flat top region before Li injection, we get a total erosion per shot and summing over the mentioned shot range, we can determine the Li erosion per head. 
This leads to a total eroded areal density of $4\times 10^{10} - 4.1\times 10^{13}$ $  \mathrm{\#/cm^{2}}$. Assuming full Li density, an areal density of $10^{15}$ $\#$/cm$^{2}$ corresponds to $0.2175$ nm. The erosion thickness of the in-situ Li coating is less than 1 nm across all shots during the 1.2 - 1.9 s time frame. Therefore, it can be inferred that all lithium reaching the samples was effectively deposited.

We can compare the erosion measurements obtained through MDS with those anticipated by the standard sheath theory during the pre-Li injection phase of the plasma flat top \cite{stangeby2000plasma}. Considering that the ion impact energy is $\mathrm{E_{i}\simeq5T_{e}}$ and $\mathrm{T_{e}\simeq30}$ eV, the expected sputtering yield of D on Li is, $\mathrm{Y_{D\rightarrow Li}}$=0.04 \cite{behrisch2007sputtering}, the expected Li erosion is 2.5 nm per shot. Although the expected erosion is low, 2.5 nm, it is higher than the one measured by MDS. This discrepancy can be attributed to the fact that the MDS spot, as mentioned, with a viewing diameter of 2.5 cm, is centrally located on the DiMES head, which is composed of graphite. As a result, the measurement of the Li signal originates predominantly from the graphite substrate rather than the mounted samples. Li intercalation with the graphite head also leads to a diminished sputtering yield. Furthermore, both the in-situ and pre-lithiated Li coatings are anticipated to be contaminated with carbon and oxygen, which has two consequences: leads to Li surface dilution and reduces sputtering yield \cite{abrams2014erosion}. The surface dilution is an effect that is anticipated to manifest itself as reduced erosion rates of any material. The discussion and methodology are consistent with those delineated in the recent study on Si injection \cite{effenberg2023situ}.

    \section{Sample composition and deuterium retention analysis}

\definecolor{CW}{RGB}{148,103,189}
\definecolor{SSLi}{RGB}{230,130,0}
\definecolor{SS}{RGB}{0,176,240}

 \begin{table*}[!htp]
		\centering
		\begin{tabular}{c|c|cc||c|c||c|c|c|c}
        \multirow{2}{*}{Head} & \multicolumn{3}{c||}{\multirow{2}{*}{Sample}} &
		 \multicolumn{2}{c||}{[nm]}  &  \multicolumn{4}{c}{[$10^{15}$ at/cm$^{2}$]}\\ 
         
            & \multicolumn{3}{c||}{} &  $\Delta$ Li& total Li  & Li  & D  &  O & C \\ \hline \hline
         1 & \multirow{2}{*}{\textbf{\textcolor{SS}{SS}}}&  \multicolumn{2}{c||}{\textbf{\textcolor{SS}{1   }}\small{(spot1, see Figure \ref{fig:SS_two_phases})}} &  260 & 260& 1185&72 &1731 &745 \\ 
         
         & & \multicolumn{2}{c||}{\textbf{\textcolor{SS}{2}}}  & \textcolor{gray}{N/A}& \textcolor{gray}{N/A}& \textcolor{gray}{N/A}&\textcolor{gray}{28} & \textcolor{gray}{N/A}& \textcolor{gray}{N/A}\\ \cline{2-10} 
         
       without  & \multirow{2}{*}{\textbf{\textcolor{SSLi}{SS+Li}}} & \multicolumn{2}{c||}{\textbf{\textcolor{SSLi}{$\mathrm{Li_{ini}=350 \ nm}$}}}  & 200& 550& 2526&23&  3717&1819 \\ 
       
        & & \multicolumn{2}{c||}{\textbf{\textcolor{SSLi}{$\mathrm{Li_{ini}=750 \ nm}$}}} & \textcolor{gray}{N/A}& \textcolor{gray}{N/A}& \textcolor{gray}{N/A}&\textcolor{gray}{127} & \textcolor{gray}{N/A}& \textcolor{gray}{N/A}\\ \cline{2-10} 
        
        RMP  & \multirow{2}{*}{\textbf{\textcolor{CW}{C+W}}} & \multirow{2}{*}{\textbf{\textcolor{CW}{$\mathrm{W=1600 \ nm}$}}} & \textbf{\textcolor{CW}{1}} & 45 & 45 & 206 & 114& 5632& 3506 \\ 
        & & & \textbf{\textcolor{CW}{2}} & 94& 94&430 &512 &3458 &2961 \\ \hline \hline

          2  & \multirow{2}{*}{\textbf{\textcolor{SS}{SS}}}&  \multicolumn{2}{c||}{\textbf{\textcolor{SS}{1}}} &  0 & 0& 0&0 &0 &0 \\ 
         
         & & \multicolumn{2}{c||}{\textbf{\textcolor{SS}{2}}}  & 0& 0& 0&0& 0& 0\\ \cline{2-10} 
         
       with  & \multirow{2}{*}{\textbf{\textcolor{SSLi}{SS+Li}}} & \multicolumn{2}{c||}{\textbf{\textcolor{SSLi}{$\mathrm{Li_{ini}=350 \ nm}$}}}  & 83& 433& 1994&25&  2799&953 \\ 
       
        & & \multicolumn{2}{c||}{\textbf{\textcolor{SSLi}{$\mathrm{Li_{ini}=750 \ nm}$}}} &36 &786 &3616 &31&5223 &2092 \\ \cline{2-10} 
        
        RMP  & \multirow{3}{*}{\textbf{\textcolor{CW}{C+W}}} & \multicolumn{2}{c||}{\textbf{\textcolor{CW}{$\mathrm{W=1000 \ nm}$}}}& 74&74 & 365& 113 & 3815& 2777 \\ 
        
        & & \multirow{2}{*}{\textbf{\textcolor{CW}{$\mathrm{W=1600 \ nm}$}}} & \textbf{\textcolor{CW}{1}} & 89&89 &441 &669 &6163 &5059 \\ 
        & & & \textbf{\textcolor{CW}{2}} & 52 & 52 & 237& 682& 4530& 3405 \\ 
          
		\end{tabular}
		\caption{Sample composition after plasma exposure. The gray values were measured by TDS and the black values by IBA. Sample SS1 from head 1 was analyzed in two different spots and twice in the same spot with 3 months in between to analyze the D decay in an argon glovebox. Here we show the first measurement from the spot with Li.}
		\label{tab:Li_thickness}
	\end{table*}

\begin{figure}
    \centering
   \input{COvsLi}
    \caption{Post-mortem measured C and O areal density dependency on the Li areal density. Filled markers represent O areal density, while unfilled markers represent C areal density. Triangles refer to reference samples, not subjected to plasma exposure, and circles refer to samples that were exposed to DIII-D plasmas. 
    }
    \label{fig:LivsCO}
\end{figure}
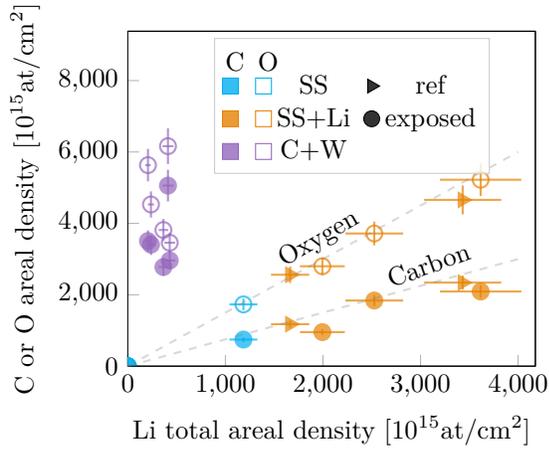

\begin{figure}
    \centering
    \resizebox{1\columnwidth}{!}{
    \input{SSLi1_2phases}
    }
    \caption{Visible camera images of sample SS1 from head showing localized Li deposition. (a) Picture after samples extraction from DIII-D. (b) and (c) Grayscale images of the same sample when inside the IB chamber. The 'x' label that denotes the location of ion beam (IB) spot. (b) image illustrates the region where IB measurement yielded 260 nm of Li, named spot1, whereas (c) depicts the IB measurement location resulting in 0 nm of Li, named spot2. }
    \label{fig:SS_two_phases}
\end{figure}
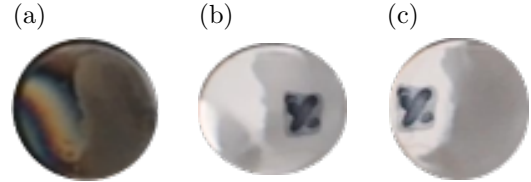

\begin{figure}
    \centering
    \input{DvsLi}
    \caption{ Post-mortem measured D areal density in all the samples vs Li areal density. }

    \label{fig:LivsD}
\end{figure}
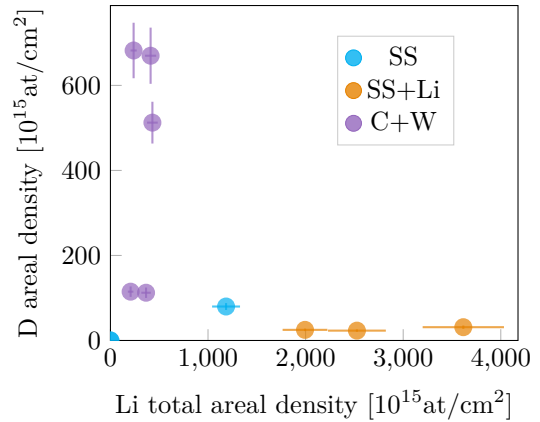

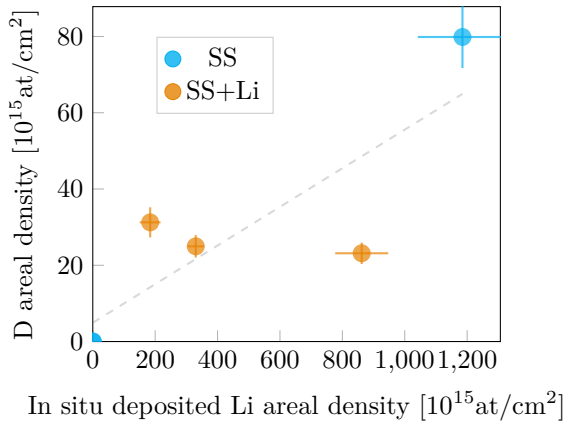
\begin{figure}
    \centering
    \input{DvsLideposited}
    \caption{Post-mortem measured D areal density in vs in-situ deposited Li areal density for the samples where D retention was dominated by Li. } 
    \label{fig:LidepositedvsD}
\end{figure}

For plasma-exposed C+W samples, carbon deposition on the sample was defined as the measured carbon areal density minus the carbon areal density of non-exposed C+W samples. The W-coated graphite samples exhibited a carbon areal density up to double that of the SS and SS+Li samples, see Figure \ref{fig:LivsCO}. Given the roughness of these samples, it is not surprising that more carbon was deposited. The grooves in the sample are likely to be filled with deposited carbon that is better protected against erosion.

The lithium content in the C+W samples was determined solely by NRA. For other sample types (SS and SS+Li), the Li content was determined by both NRA and EBS, as detailed in \cite{morbey2024deuterium}. To quantify the amount of Li deposited during the injection of Li into the plasma of the SS+Li samples, we calculated the difference between the Li areal density of the exposed samples and that of the non-exposed reference SS+Li samples. From the composition analysis, it is evident that the Li coating from the injection of Li into the plasma was uneven within the same head, see Table \ref{tab:Li_thickness}. For head 1 the thickness of Li deposited on the different samples varied between 45-260 nm. The thickness was calculated by converting the areal density to an average thickness of pure Li  and assuming full density of the coating. Moreover, in Figure \ref{fig:SS_two_phases} is shown that sample SS1 from head 1 had Li deposition only on half of its surface: this highlights the nonuniform Li coating and may help to explain the large variation between samples. This was the sample where the surface non-uniformity was most noticeable, and the only one where two IB spots were measured. For head 2, where RMPs were used, no Li deposition was detected in either of the stainless steel samples. The rest of the samples in head 2 showed a deposited Li coating between 36-90 nm. Air contamination of the Li layers is observed in Figure \ref{fig:LivsCO} where the areal density of both impurities C and O increases linearly with Li areal density and even the non-plasma exposed samples align with respect to the Li:C:O ratio.

For deuterium retention, despite efforts to minimize desorption caused by chemical interactions with air, a significant portion of deuterium is expected to have desorbed prior to the IBA.
To study this D desorption in an Ar atmosphere, sample SS1 was measured with IBA 3 months and 6 months after plasma exposure. During the 3 months between measurements, the sample was stored in an Ar glove box. The areal density of Li, O and C did not change, however, the areal density of D was reduced by 63$\%$, from (79 $\pm$ 7) to (29 $\pm$ 3) $\times \mathrm{10^{15}}$ $\mathrm{at/cm^{2}}$.
Moreover, comparing TDS results with IBA results is nontrivial, as TDS was performed within less than 1 week from the experiment date and IBA was performed 3 months later. However, we can still compare the retention of D in the different IB measurements, this is shown in Figure \ref{fig:LivsD}. This figure displays all the IB analyzed samples, indicating the set to which each sample belongs using distinct markers. It is seen that D retention does not correlate with the Li areal density. 
Specifically, within holder 1, the Li concentration after plasma exposure in the SS+Li sample is approximately twice that of the SS sample. However, the D concentration is higher in the SS sample compared to the SS+Li sample. The comparison of the pre-lithiated samples of head 2 to each other corrobates the observations made with samples from head 1: retention of D does not depend on the thickness of the pre-deposited Li coating. In Figure \ref{fig:LivsD} we also see that the W coated C samples have a higher D retention, despite having a relative low Li deposition. As seen in Figure \ref{fig:LivsCO}, the C areal density of these samples surpasses that of Li in these samples. Due to the strong affinity of carbon for hydrogen isotopes, deuterium retention in these samples is primarily attributed to carbon rather than lithium. Consequently, deuterium areal density should not be included in studies examining the relationship between lithium areal density and deuterium retention.
In Figure \ref{fig:LidepositedvsD}, excluding the W-coated samples, a weak correlation between the deuterium areal density and the in-situ deposited lithium areal density is observed, suggesting that retention is driven by co-deposition.

    \section{Discussion}\label{sec:Discussion}
    From the analysis of pre-lithiated samples with varying thicknesses, we conclude that D retention in surfaces coated with solid Li exposed to far scrape-off-layer D plasmas is minimal compared to retention in Li-D co-deposits and is independent of the film's thickness. It is important to note that these findings are only applicable to solid Li films. Although no direct temperature measurements were conducted during our experiments, prior DiMES exposures under similar plasma conditions suggest that sample temperatures remained below 100 $^{\circ}$C, staying well below the Li melting point.

These results indicate that D interacts with Li at the surface, likely forming lithium deuteride (LiD). The formation of a LiD layer, combined with a solid Li substrate, prevents D from diffusing deeper into the material, thereby limiting retention. This conclusion aligns with the findings reported in \cite{ou2022deuterium}.

In contrast, Li injection via IPD demonstrates that Li-D co-deposits are the main driver of D retention at low temperatures. This conclusion is supported by the observation that fuel retention levels were similar in pre-lithiated and non-pre-lithiated samples exposed to identical D plasma conditions with in-situ Li injection. Furthermore, the retention observed in SS and SS+Li samples is attributed to Li rather than carbon from the walls, as the Li content exceeds the carbon content. Moreover, Figure \ref{fig:LivsCO} shows that the C and O content of these samples increases linearly with Li, indicating that the observed C is primarily due to contamination rather than deposition from the walls. Additionally, the D:C ratio is significantly lower than that observed in C+W samples, where the carbon content outweighs that of lithium.

In the different scenarios where fuel retention in Li was studied, it was observed that the use of RMPs led to less Li deposition on the samples, despite a 28$\%$ increase in Li injection compared to non-RMP conditions. This reduction in deposition is attributed to RMPs enhancing outward plasma transport, which increases D flux and erosion \cite{evans2006edge, hager2019gyrokinetic}. It should be noted, this effect could go unnoticed by Langmuir probes, as they are positioned at different toroidal locations from DiMES.

In the second set of samples, both polished stainless steel surfaces placed on opposite sides of the holder showed no detectable deposition of Li or C. This suggests that thin Li layers are less likely to adhere to polished metal surfaces, possibly due to a lower surface sticking coefficient or reduced trapping efficiency compared to rougher samples containing C and O impurities.

With respect to the choice of samples, the various substrates used in this experiment presented different degrees of complexity during the ion beam analysis. For the C+W samples because of the surface roughness and sample changes, more effects needed to be taken into account during the analysis. 
	
    \section{Conclusions}\label{sec:conclusions}
    In this work, we exposed SS, SS with pre-deposited Li, and C with W-coating samples to DIII-D H-mode plasma using DiMES. The samples were introduced in the far SOL, and Li was injected into the plasmas with the Impurity Powder Dropper. We studied Li erosion by D, uniformity of the Li coatings, fuel retention dependence on pre-deposited Li thickness, and the comparison of fuel retention between pre-deposited Li coatings and Li-D co-deposits. The thickness of the in-situ deposited Li layer varied between 45-260 nm and for scenarios with RMPs between 0-90 nm, although similar Li mass injections were conducted.
    We conclude that at temperatures below the melting point of lithium, the retention of deuterium is independent of the thickness of the pre-deposited lithium at least up to the Li layer thickness studied here. Moreover, Li-D co-deposition is the primary driver for fuel retention, although the correlation is not strong in these studies.
    With regard to in-situ diagnostics, at the far SOL, pre-deposited Li and Li deposited in-situ showed lower erosion than the expected value from the sputtering yield of Li.
    
    D content after three months, with the sample stored in an Ar glove box, emphasizes the necessity of minimizing air contamination and conducting in-situ measurements to ensure accurate assessment of D retention during plasma exposure.
    The thickness of the Li coating was observed to be non-uniform through the same head and seemed to vary with the surface morphology where Li deposits. This is highlighted by the polished SS samples of the second set of samples, where no Li was observed, while other samples had a Li deposition of 36-90 nm. To achieve lower recycling regimes in future fusion reactors that do not include the use of liquid metals, operando Li coating by impurity powder dropper is a viable option. From this work, it is expected that Li injection during plasma operation is a more efficient method for fuel pumping than pre-lithiated walls, at least at temperatures below the melting point of Li. Moreover, for fusion reactor designs that incorporate liquid lithium divertors \cite{goldston2016lithium} due to their ability to handle high heat fluxes, concerns arise regarding fuel retention from Li-D co-deposits forming on the first wall.

    \section*{Acknowledgments}

This material is based upon work supported by the U.S. Department of Energy, Office of Science, Office of Fusion Energy Sciences, using the DIII-D National Fusion Facility, a DOE Office of Science user facility, under Awards DE-AC02-09CH11466, DE-FC02-04ER54698, DE-NA0003525,  DE-SC0019256, DE-SC0023378, DE-SC0015877 and DE-AC52-07NA27344. This work has been carried out within the framework of the EUROfusion Consortium, funded by the European Union via the Euratom Research and Training Programme (Grant Agreement No 101052200 EUROfusion). Views and opinions expressed are however those of the author(s) only and do not necessarily reflect those of the European Union or the European Commission. Neither the European Union nor the European Commission can be held responsible for them.  This work is part of the research programme "The Leidenfrost divertor: a lithium vapor shield for extreme heat loads to fusion reactor walls" with project number VI.Vidi.198.018, which is (partly) financed by NWO. The DIII-D data shown in this paper can be obtained in digital format by following the links at \url{https://fusion.gat.com/global/D3D_DMP}. The United States Government retains a non-exclusive, paid-up, irrevocable, world-wide license to publish or reproduce the published form of this manuscript, or allow others to do so, for United States Government purposes.

\textbf{Disclaimer:} This report was prepared as an account of work sponsored by an agency of the United States Government. Neither the United States Government nor any agency thereof, nor any of their employees, makes any warranty, express or implied, or assumes any legal liability or responsibility for the accuracy, completeness, or usefulness of any information, apparatus, product, or process disclosed, or represents that its use would not infringe privately owned rights. Reference herein to any specific commercial product, process, or service by trade name, trademark, manufacturer, or otherwise does not necessarily constitute or imply its endorsement, recommendation, or favoring by the United States Government or any agency thereof. The views and opinions of authors expressed herein do not necessarily state or reflect those of the United States Government or any agency thereof.
	\section*{References}
	\bibliographystyle{unsrt}
	\bibliography{bibliography}

\end{document}

%% file: Pre-plasma-figures.tex
{\scalefont{0.9}
\definecolor{m}{RGB}{191,0,191}
\definecolor{yellow}{RGB}{255,165,0}
\definecolor{greenm}{RGB}{0,128,0}
\definecolor{pink}{RGB}{255,105,180}
\definecolor{babyblue}{RGB}{0,191,255}
\definecolor{darkblue}{RGB}{0,0,255}
\definecolor{purple}{RGB}{128,0,128}

\begin{tikzpicture}

\node[above left, inner sep=0] (image) at (-3,0){\includegraphics[trim=180 50 230 20, clip,width=3.5cm]{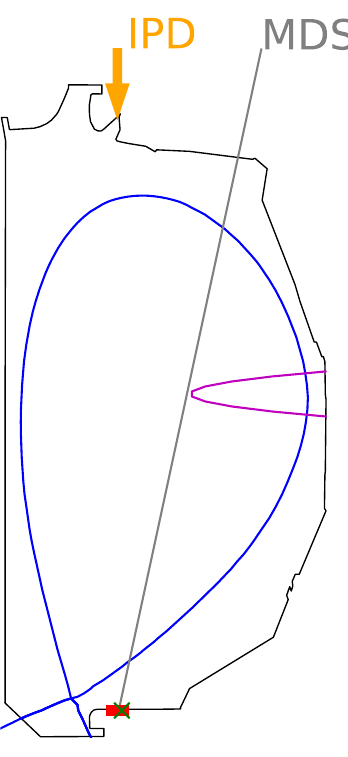}};
         \draw[stealth-, ultra thick,red] (-5.3,0.35) -- (-3.4,0.0)
           node[red, fill=white]{DiMES 150$^\circ$};  
            \draw[stealth-,  thick,black] (-6.8,1) -- (-6.8,0);
             \draw[stealth-,  thick,black] (-5.8,0) -- (-6.8,0);
             \node[black] at (-7,0.5) {Z};
             \node[black] at (-6.3,-0.2) {R};

        \draw[stealth-, ultra thick,greenm] (-5.15,0.5) -- (-3.4,0.4)
           node[greenm, fill=white]{LP 180$^\circ$};  
           
         \node (a) [left,m,fill=white] at (-1.75,3.7) { SPRED    };

         \draw[draw=white, fill=white] (-5.8,7) rectangle ++(3.4,0.5);
         \node (a) [left,gray,fill=white] at (-3,7.5) { MDS};
         \node (a) [left,gray,fill=white] at (-3,7.1) {L5 150$^\circ$ };
         \node (a) [left,gray] at (-3.4,6.57) {L6};
         \node (a) [left,gray] at (-4,6.57) {L4};
        \draw[thick,gray] (-4.96,0.5) -- (-3.93,6.87);
        \draw[thick,gray] (-5.73,0.19) -- (-3.93,6.87);

         \node (a) [left,yellow,fill=white] at (-4.5,7.4) { IPD  195$^\circ$  };

       \node [inner sep=0pt ](a) at (0,5.8) 
         {\includegraphics[trim=1 1 1 1, clip, width=3.4cm]{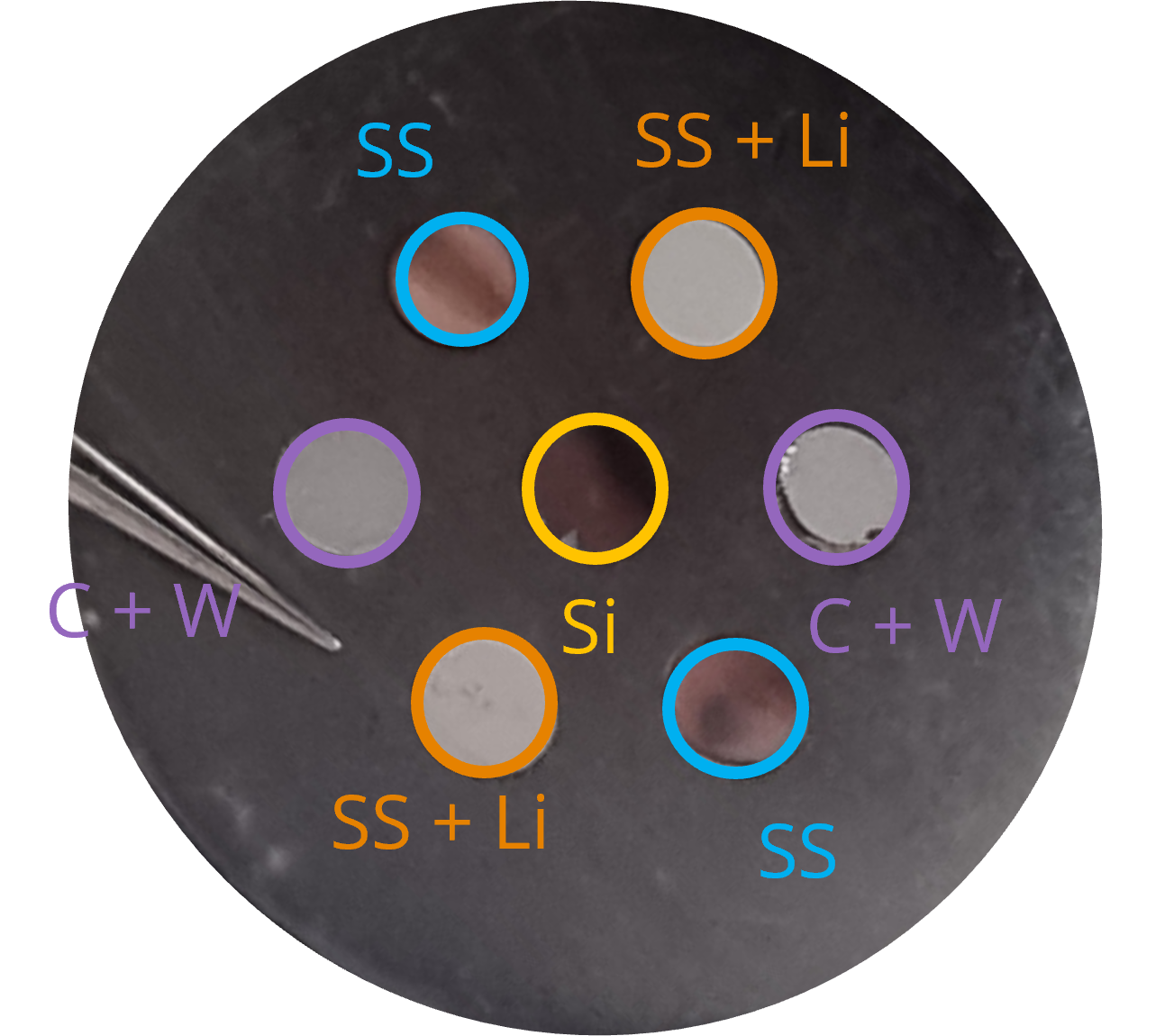}};
        \node [inner sep=0pt ](b) at (0,1.7) 
         {\includegraphics[trim=1 1 1 1, clip, width=3.1cm]{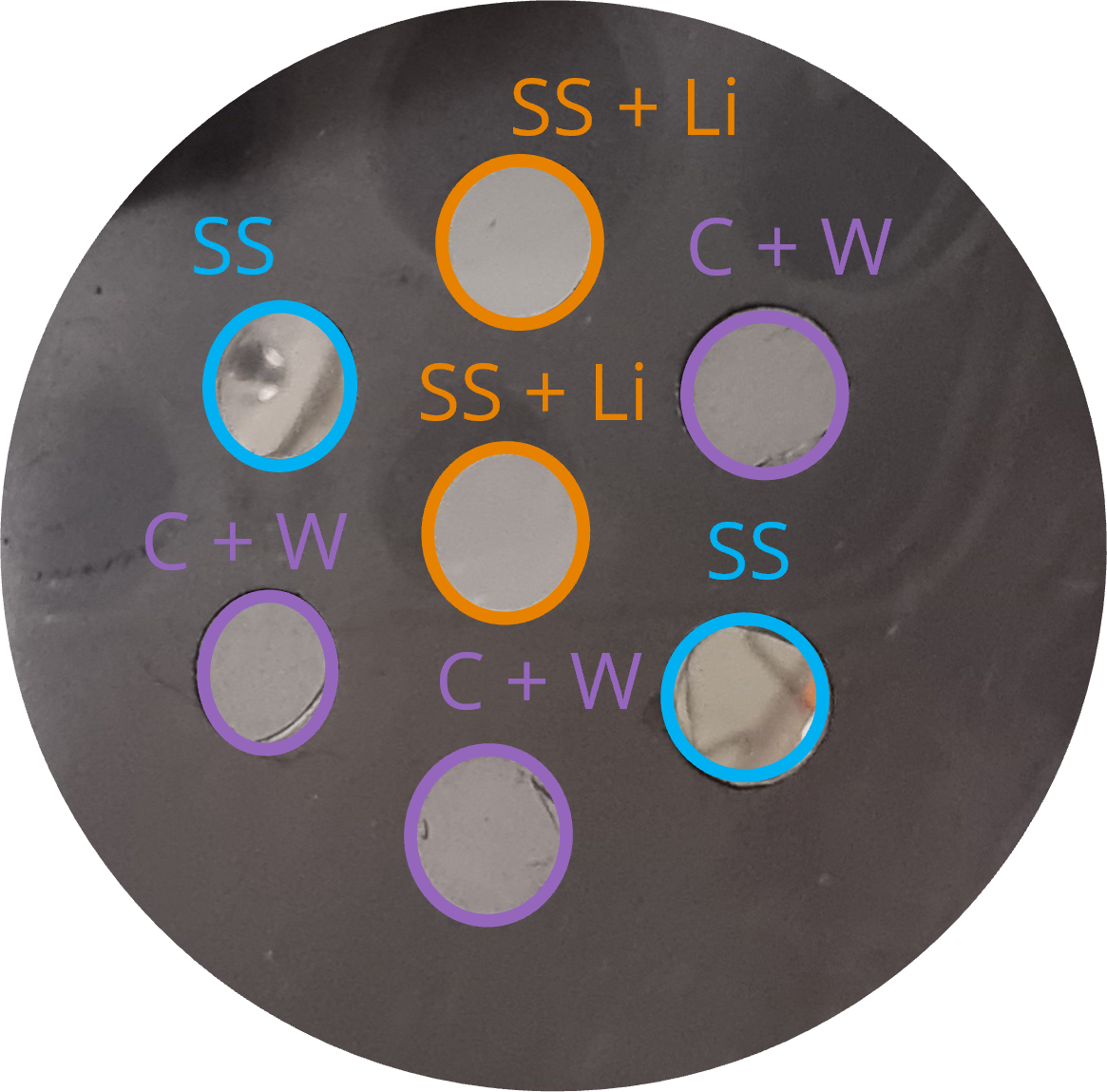}};
      
         \node (aa) [left,black] at (-6.5,7) { (a)    };
          \node (aa) [left,black] at (-1.5,7) { (b)    };
         \node (bb) [left,black] at (-1.5,3) {(c)   };

 \draw[thick] (-1.3,4) -- (1.3,4); 
    \draw[thick] (-1.3,4.2) -- (-1.3,3.8); 
    \draw[thick] (1.3,4.2) -- (1.3,3.8); 
    
    \node at (0, 3.7) {4.8 cm}; 

    \draw[thick] (-6.45,-0.4) -- (-3.24,-0.4); 
    \draw[thick] (-6.45,-0.6) -- (-6.45,-0.2); 
    \draw[thick] (-3.24,-0.6) -- (-3.24,-0.2); 
    
    \node at (-5, -0.55) {130 cm}; 
\end{tikzpicture}
}

%% file: plasma_parameters.tex
{\scalefont{0.95}
\begin{tikzpicture}
\definecolor{pinkm}{RGB}{255,105,180}
\definecolor{bluem}{RGB}{41,41,255}
\definecolor{deepskyblue0176240}{RGB}{0,176,240}

    \node [
        above right,
        inner sep=0] (image) at (0,0) {\includegraphics[trim= 40 40 43 55, clip, width=0.9\columnwidth, height=8cm]{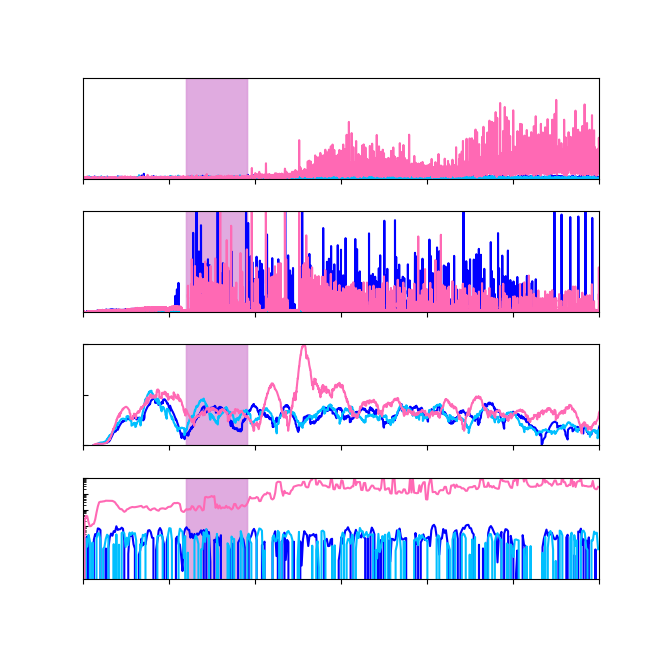}};
    \begin{scope}[
    x={($0.1*(image.south east)$)},
    y={($0.1*(image.north west)$)}]
     
     
      
        \draw[] (0.21,0.4) 
           node[black, fill=white]{0}; 
        \draw[] (0.21,2.3) 
           node[black]{$10^{15}$}; 

        \draw[] (0.21,2.9) 
           node[black, fill=white]{0}; 
        \draw[] (0.21,4.8) 
           node[black]{$10^{18}$}; 

        \draw[] (0.21,5.5) 
           node[black, fill=white]{0}; 
        \draw[] (0.21,7.3) 
           node[black, fill=white]{1}; 

       
          \draw[] (0.21,8) 
           node[black, fill=white]{0}; 
        \draw[] (0.21,9.9) 
           node[black, fill=white]{1};

       \draw[stealth-, ultra thick,white] (-0.2,8.75*1.05) 
           node[black, fill=white, rotate=90]{ Li III [a.u.]};
       \draw[stealth-, ultra thick,white] (-0.2,6.4) 
           node[black, fill=white, rotate=90]{ D$_\alpha$[a.u.]};
        \draw[stealth-, ultra thick,white] (-0.2,3.5) 
           node[black, fill=white, rotate=90]{$\Gamma_{\mathrm{D^{+}}}$[$ \mathrm{\#/cm^{2}s}$]};
       \draw[stealth-, ultra thick,white] (-0.2,0.7) 
           node[black, fill=white, rotate=90]{  $\Gamma_{\mathrm{Li^{+}}}$[$\mathrm{\#/cm^{2}s}$]};
     \draw[stealth-, ultra thick,white] (0.53,-0.2) 
           node[black, fill=white]{ 0}; 
            \draw[stealth-, ultra thick,white] (2.09,-0.2) 
           node[black, fill=white]{ 1}; 
            \draw[stealth-, ultra thick,white] (3.65,-0.2) 
           node[black, fill=white]{ 2}; 
            \draw[stealth-, ultra thick,white] (5.2,-0.2) 
           node[black, fill=white]{ 3}; 
              \draw[stealth-, ultra thick,white] (6.8,-0.2)  
           node[black, fill=white]{ 4}; 
              \draw[stealth-, ultra thick,white] (8.3,-0.2) 
           node[black, fill=white]{ 5}; 
              \draw[stealth-, ultra thick,white] (9.85,-0.2) 
           node[black, fill=white]{ 6}; 

          \node[black] at (5,-0.8)  { Time [s]}; 

           \draw[stealth-, ultra thick,white] (1,9.5) 
           node[black, fill=white]{ (a)}; 
           \draw[stealth-, ultra thick,white] (1,6.9) 
           node[black, fill=white]{ (b)}; 
           \draw[stealth-, ultra thick,white] (1,4.2) 
           node[black, fill=white]{ (c)}; 
       \draw[stealth-, ultra thick] (1,1.6) 
           node[black, fill=white]{ (d)}; 

            \draw[stealth-, ultra thick,white] (3,10.6) 
           node[bluem]{\textbf{ref DiMES retracted $\#$ 196020}}; 
           \draw[stealth-, ultra thick,white] (3,10.2) 
           node[deepskyblue0176240]{\textbf{ref DiMES inserted $\#$ 196021}}; 
           \draw[stealth-, ultra thick,white] (8,10.2) 
           node[pinkm]{\textbf{Li inj. $\#$ 196024}}; 
        
    \end{scope}
\end{tikzpicture}
}

%% file: COvsLi.tex
\begin{tikzpicture}
\usetikzlibrary{shapes.geometric}

\definecolor{darkgray176}{RGB}{176,176,176}
\definecolor{darkorange2301300}{RGB}{230,130,0}
\definecolor{deepskyblue0176240}{RGB}{0,176,240}
\definecolor{lightgray204}{RGB}{204,204,204}
\definecolor{mediumpurple148103189}{RGB}{148,103,189}

\begin{axis}[
legend cell align={left},
legend columns=2,
legend style={
  fill opacity=0.8,
  draw opacity=1,
  text opacity=1,
  at={(0.5,0.91)},
  anchor=north,
  draw=lightgray204
},
tick align=inside,
clip marker paths=true,
tick pos=left,
x grid style={darkgray176},
xlabel={Li total areal density \(\displaystyle [\mathrm{10^{15}at/cm^{2}]}\)},
xmin=0, xmax=4176.53,
xtick style={color=darkgray176},
y grid style={darkgray176},
ylabel={C or O areal density \(\displaystyle [\mathrm{10^{15}at/cm^{2}]}\)},
ymin=0, ymax=9382.05323388,
ytick style={color=darkgray176},
tick align=inside,
tick pos=left,
clip mode=individual,
]
\path [draw=mediumpurple148103189, draw opacity=0.7, semithick]
(axis cs:375.215330350027,2961.43780604)
--(axis cs:484.984669649973,2961.43780604);

\path [draw=mediumpurple148103189, draw opacity=0.7, semithick]
(axis cs:430.1,2706.07183961425)
--(axis cs:430.1,3216.80377246575);

\path [draw=mediumpurple148103189, draw opacity=0.7, semithick]
(axis cs:375.215330350027,3458)
--(axis cs:484.984669649973,3458);

\path [draw=mediumpurple148103189, draw opacity=0.7, semithick]
(axis cs:430.1,3178.81856425487)
--(axis cs:430.1,3737.18143574513);

\path [draw=mediumpurple148103189, draw opacity=0.7, semithick]
(axis cs:178.302230992387,3506)
--(axis cs:235.497769007613,3506);

\path [draw=mediumpurple148103189, draw opacity=0.7, semithick]
(axis cs:206.9,3206.44901867263)
--(axis cs:206.9,3805.55098132737);

\path [draw=mediumpurple148103189, draw opacity=0.7, semithick]
(axis cs:178.302230992387,5632)
--(axis cs:235.497769007613,5632);

\path [draw=mediumpurple148103189, draw opacity=0.7, semithick]
(axis cs:206.9,5176.45645097166)
--(axis cs:206.9,6087.54354902834);

\path [draw=darkorange2301300, draw opacity=0.7, semithick]
(axis cs:2229.90576113399,1839)
--(axis cs:2822.09423886601,1839);

\path [draw=darkorange2301300, draw opacity=0.7, semithick]
(axis cs:2526,1680.70149784417)
--(axis cs:2526,1997.29850215583);

\path [draw=darkorange2301300, draw opacity=0.7, semithick]
(axis cs:2229.90576113399,3717)
--(axis cs:2822.09423886601,3717);

\path [draw=darkorange2301300, draw opacity=0.7, semithick]
(axis cs:2526,3384.43733806281)
--(axis cs:2526,4049.56266193719);

\path [draw=deepskyblue0176240, draw opacity=0.7, semithick]
(axis cs:1041.85054049562,745.6)
--(axis cs:1328.14945950438,745.6);

\path [draw=deepskyblue0176240, draw opacity=0.7, semithick]
(axis cs:1185,679.258677728649)
--(axis cs:1185,811.941322271351);

\path [draw=deepskyblue0176240, draw opacity=0.7, semithick]
(axis cs:1041.85054049562,1731)
--(axis cs:1328.14945950438,1731);

\path [draw=deepskyblue0176240, draw opacity=0.7, semithick]
(axis cs:1185,1566.69600688048)
--(axis cs:1185,1895.30399311952);

\path [draw=mediumpurple148103189, draw opacity=0.7, semithick]
(axis cs:205.838752276016,3404.699619464)
--(axis cs:269.361247723984,3404.699619464);

\path [draw=mediumpurple148103189, draw opacity=0.7, semithick]
(axis cs:237.6,3113.81805451293)
--(axis cs:237.6,3695.58118441507);

\path [draw=mediumpurple148103189, draw opacity=0.7, semithick]
(axis cs:205.838752276016,4530)
--(axis cs:269.361247723984,4530);

\path [draw=mediumpurple148103189, draw opacity=0.7, semithick]
(axis cs:237.6,4164.25068897212)
--(axis cs:237.6,4895.74931102788);

\path [draw=mediumpurple148103189, draw opacity=0.7, semithick]
(axis cs:354.51290045662,5058.7373656)
--(axis cs:468.88709954338,5058.7373656);

\path [draw=mediumpurple148103189, draw opacity=0.7, semithick]
(axis cs:411.7,4615.57825172038)
--(axis cs:411.7,5501.89647947962);

\path [draw=mediumpurple148103189, draw opacity=0.7, semithick]
(axis cs:354.51290045662,6163)
--(axis cs:468.88709954338,6163);

\path [draw=mediumpurple148103189, draw opacity=0.7, semithick]
(axis cs:411.7,5662.30502630951)
--(axis cs:411.7,6663.69497369049);

\path [draw=mediumpurple148103189, draw opacity=0.7, semithick]
(axis cs:316.524369077286,2777)
--(axis cs:413.675630922714,2777);

\path [draw=mediumpurple148103189, draw opacity=0.7, semithick]
(axis cs:365.1,2534.36584365755)
--(axis cs:365.1,3019.63415634245);

\path [draw=mediumpurple148103189, draw opacity=0.7, semithick]
(axis cs:316.524369077286,3815)
--(axis cs:413.675630922714,3815);

\path [draw=mediumpurple148103189, draw opacity=0.7, semithick]
(axis cs:365.1,3505.67722040978)
--(axis cs:365.1,4124.32277959022);

\path [draw=deepskyblue0176240, draw opacity=0.7, semithick]
;

\path [draw=deepskyblue0176240, draw opacity=0.7, semithick]
;

\path [draw=deepskyblue0176240, draw opacity=0.7, semithick]
;

\path [draw=deepskyblue0176240, draw opacity=0.7, semithick]
;

\path [draw=deepskyblue0176240, draw opacity=0.7, semithick]
;

\path [draw=deepskyblue0176240, draw opacity=0.7, semithick]
;

\path [draw=deepskyblue0176240, draw opacity=0.7, semithick]
;

\path [draw=deepskyblue0176240, draw opacity=0.7, semithick]
;

\path [draw=darkorange2301300, draw opacity=0.7, semithick]
(axis cs:1764.3215024548,953.8)
--(axis cs:2223.6784975452,953.8);

\path [draw=darkorange2301300, draw opacity=0.7, semithick]
(axis cs:1994,873.276489680668)
--(axis cs:1994,1034.32351031933);

\path [draw=darkorange2301300, draw opacity=0.7, semithick]
(axis cs:1764.3215024548,2799)
--(axis cs:2223.6784975452,2799);

\path [draw=darkorange2301300, draw opacity=0.7, semithick]
(axis cs:1994,2551.88793054506)
--(axis cs:1994,3046.11206945494);

\path [draw=darkorange2301300, draw opacity=0.7, semithick]
(axis cs:3199.29482712257,2092)
--(axis cs:4032.70517287743,2092);

\path [draw=darkorange2301300, draw opacity=0.7, semithick]
(axis cs:3616,1915.30563564412)
--(axis cs:3616,2268.69436435588);

\path [draw=darkorange2301300, draw opacity=0.7, semithick]
(axis cs:3199.29482712257,5223)
--(axis cs:4032.70517287743,5223);

\path [draw=darkorange2301300, draw opacity=0.7, semithick]
(axis cs:3616,4764.77016348313)
--(axis cs:3616,5681.22983651687);

\path [draw=darkorange2301300, draw opacity=0.7, semithick]
(axis cs:1467.75444774902,1177)
--(axis cs:1860.24555225098,1177);

\path [draw=darkorange2301300, draw opacity=0.7, semithick]
(axis cs:1664,1075.2492649268)
--(axis cs:1664,1278.7507350732);

\path [draw=darkorange2301300, draw opacity=0.7, semithick]
(axis cs:3036.90160296073,2341)
--(axis cs:3827.09839703927,2341);

\path [draw=darkorange2301300, draw opacity=0.7, semithick]
(axis cs:3432,2143.21281861974)
--(axis cs:3432,2538.78718138026);

\path [draw=darkorange2301300, draw opacity=0.7, semithick]
(axis cs:1467.75444774902,2561)
--(axis cs:1860.24555225098,2561);

\path [draw=darkorange2301300, draw opacity=0.7, semithick]
(axis cs:1664,2329.5079345331)
--(axis cs:1664,2792.4920654669);

\path [draw=darkorange2301300, draw opacity=0.7, semithick]
(axis cs:3036.90160296073,4658)
--(axis cs:3827.09839703927,4658);

\path [draw=darkorange2301300, draw opacity=0.7, semithick]
(axis cs:3432,4252.9992058517)
--(axis cs:3432,5063.0007941483);

\addplot [semithick, mediumpurple148103189, opacity=0.7, mark=*, mark size=3, mark options={solid}, only marks]
table {%
430.1 2961.43780604
};
\addplot [semithick, mediumpurple148103189, opacity=0.7, mark=o, mark size=3, mark options={solid,fill opacity=0}, only marks]
table {%
430.1 3458
};
\addplot [semithick, mediumpurple148103189, opacity=0.7, mark=*, mark size=3, mark options={solid}, only marks]
table {%
206.9 3506
};
\addplot [semithick, mediumpurple148103189, opacity=0.7, mark=o, mark size=3, mark options={solid,fill opacity=0}, only marks]
table {%
206.9 5632
};
\addplot [semithick, darkorange2301300, opacity=0.7, mark=*, mark size=3, mark options={solid}, only marks]
table {%
2526 1839
};
\addplot [semithick, darkorange2301300, opacity=0.7, mark=o, mark size=3, mark options={solid,fill opacity=0}, only marks]
table {%
2526 3717
};
\addplot [semithick, deepskyblue0176240, opacity=0.7, mark=*, mark size=3, mark options={solid}, only marks]
table {%
1185 745.6
};
\addplot [semithick, deepskyblue0176240, opacity=0.7, mark=o, mark size=3, mark options={solid,fill opacity=0}, only marks]
table {%
1185 1731
};
\addplot [semithick, mediumpurple148103189, opacity=0.7, mark=*, mark size=3, mark options={solid}, only marks]
table {%
237.6 3404.699619464
};
\addplot [semithick, mediumpurple148103189, opacity=0.7, mark=o, mark size=3, mark options={solid,fill opacity=0}, only marks]
table {%
237.6 4530
};
\addplot [semithick, mediumpurple148103189, opacity=0.7, mark=*, mark size=3, mark options={solid}, only marks]
table {%
411.7 5058.7373656
};
\addplot [semithick, mediumpurple148103189, opacity=0.7, mark=o, mark size=3, mark options={solid,fill opacity=0}, only marks]
table {%
411.7 6163
};
\addplot [semithick, mediumpurple148103189, opacity=0.7, mark=*, mark size=3, mark options={solid}, only marks]
table {%
365.1 2777
};
\addplot [semithick, mediumpurple148103189, opacity=0.7, mark=o, mark size=3, mark options={solid,fill opacity=0}, only marks]
table {%
365.1 3815
};
\addplot [semithick, deepskyblue0176240, opacity=0.7, mark=*, mark size=3, mark options={solid}, only marks]
table {%
0 0
};
\addplot [semithick, deepskyblue0176240, opacity=0.7, mark=o, mark size=3, mark options={solid,fill opacity=0}, only marks]
table {%
0 0
};
\addplot [semithick, deepskyblue0176240, opacity=0.7, mark=*, mark size=3, mark options={solid}, only marks]
table {%
0 0
};
\addplot [semithick, deepskyblue0176240, opacity=0.7, mark=o, mark size=3, mark options={solid,fill opacity=0}, only marks]
table {%
0 0
};
\addplot [semithick, darkorange2301300, opacity=0.7, mark=*, mark size=3, mark options={solid}, only marks]
table {%
1994 953.8
};
\addplot [semithick, darkorange2301300, opacity=0.7, mark=o, mark size=3, mark options={solid,fill opacity=0}, only marks]
table {%
1994 2799
};
\addplot [semithick, darkorange2301300, opacity=0.7, mark=*, mark size=3, mark options={solid}, only marks]
table {%
3616 2092
};
\addplot [semithick, darkorange2301300, opacity=0.7, mark=o, mark size=3, mark options={solid,fill opacity=0}, only marks]
table {%
3616 5223
};
\addplot [semithick, darkorange2301300, opacity=0.7, mark=triangle*, mark size=3, mark options={solid,rotate=270}]
table {%
1664 1177
};
\addplot [semithick, darkorange2301300, opacity=0.7, mark=triangle*, mark size=3, mark options={solid,rotate=270}]
table {%
3432 2341
};
\addplot [semithick, darkorange2301300, opacity=0.7, mark=triangle*, mark size=3, mark options={solid,rotate=270}]
table {%
1664 2561
};
\addplot [semithick, darkorange2301300, opacity=0.7, mark=triangle*, mark size=3, mark options={solid,rotate=270}]
table {%
3432 4658
};

\end{axis}
\node[text width=2cm,rotate=30] at (2.9,1.9) {Oxygen};
\node[text width=2cm,rotate=20] at (4.4,1.5) {Carbon};

\begin{axis}[
axis y line=right,
xtick=\empty,
xmin=0, xmax=4176.53,
y grid style={darkgray176},
ymin=0, ymax=9382.05323388,
ytick pos=right,
legend columns=2, 
legend style={
    /tikz/column 2/.style={
        column sep=3pt,
    },
  fill opacity=0.8,
  draw opacity=1,
  text opacity=1,
  at={(0.55,0.98)},
  anchor=north,
  draw=lightgray204
},
ytick style={color=darkgray176},
ytick align= inside,
yticklabel style={anchor=west},
clip mode=individual,
ytick=\empty,
num1/.style={deepskyblue0176240, mark= square*, only marks, mark size=3},
num2/.style={deepskyblue0176240, mark=square, mark options={solid,fill opacity=0}, only marks, mark size=3},
combo legend/.style={
    legend image code/.code={
        \draw [/pgfplots/num1] plot coordinates {(1mm,0cm)};
        \draw plot coordinates {(2.5mm,-3pt)} node { };
        \draw [/pgfplots/num2] plot coordinates {(5.4mm,0cm)};
        }
    },
num11/.style={darkorange2301300, mark= square*, only marks, mark size=3},
num12/.style={darkorange2301300, mark= square,mark options={solid,fill opacity=0}, only marks, mark size=3},
something/.style={
    legend image code/.code={
        \draw [/pgfplots/num11] plot coordinates {(1mm,0cm)};
        \draw plot coordinates {(2.5mm,-3pt)} node { };
        \draw [/pgfplots/num12] plot coordinates {(5.4mm,0cm)};
        }
    },
num111/.style={mediumpurple148103189, mark= square*, only marks, mark size=3},
num112/.style={mediumpurple148103189, mark= square,mark options={solid,fill opacity=0},  only marks, mark size=3},
something1/.style={
    legend image code/.code={
        \draw [/pgfplots/num111] plot coordinates {(1mm,0cm)};
        \draw plot coordinates {(2.5mm,-3pt)} node {\textcolor{white}{,}};
        \draw [/pgfplots/num112] plot coordinates {(5.4mm,0cm)};
        }
    }
]

\addplot [semithick, gray, dashed, opacity=0.3, forget plot]
table {%
0 0
4000 6000
};

\addplot [semithick, gray, dashed, opacity=0.3,forget plot]
table {%
0 0
4000 3000
};


    \addlegendentry{\hspace{-.6cm}\textbf{A title}}
    \addlegendimage{white, mark=triangle*, only marks, mark size=3}
    \addlegendimage{white, mark=triangle*, only marks, mark size=3}

    \addlegendimage{combo legend}
    \addlegendimage{black, mark=triangle*, mark options={solid,rotate=270}, only marks, mark size=3}

    \addlegendimage{something}
   \addlegendimage{black, mark=*, only marks, mark size=3}
    \addlegendimage{something1}

\addlegendimage{black, mark=*, only marks, mark size=3}

\legend{\textcolor{white}{a},\textcolor{white}{a},SS,ref,SS+Li, exposed,C+W}

\end{axis}
\node[text width=4cm] at (3.3,4.05) {C\textcolor{white}{a}O};
\end{tikzpicture}

%% file: SSLi1_2phases.tex
\begin{tikzpicture}

\node[] (image) at (0.5,0){\includegraphics[scale=0.5]{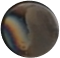}};
\node[rotate=90] at (image) {};

        \node (fig2) at (1.5,0)
       {\includegraphics[scale=0.5]{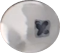}}; 
       \node (fig2) at (2.5,0)
       {\includegraphics[scale=0.5]{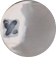}}; 
            \node[black, scale=0.4] at   (0.2,0.5){(a)};          
           \node[black,scale=0.4] at (1.2,0.5) {(b)}; 
               \node[black, scale=0.4]  at (2.2,0.5) {(c)};

     
\end{tikzpicture}

%% file: DvsLi.tex
\begin{tikzpicture}

\definecolor{darkgray176}{RGB}{176,176,176}
\definecolor{darkorange2301300}{RGB}{230,130,0}
\definecolor{deepskyblue0176240}{RGB}{0,176,240}
\definecolor{lightgray204}{RGB}{204,204,204}
\definecolor{mediumpurple148103189}{RGB}{148,103,189}

\begin{axis}[
legend columns=1,
legend style={
    /tikz/column 2/.style={
        column sep=3pt,
    },
  fill opacity=0.8,
  draw opacity=1,
  text opacity=1,
  at={(0.7,0.91)},
  anchor=north,
  draw=lightgray204
},tick align=inside,
tick pos=left,
x grid style={darkgray176},
xlabel={Li total areal density \(\displaystyle [\mathrm{10^{15}at/cm^{2}]}\)},
xmin=0, xmax=4176.48,
xtick style={color=darkgray176},
y grid style={darkgray176},
ylabel={D areal density \(\displaystyle [\mathrm{10^{15}at/cm^{2}]}\)},
ymin=0, ymax=787.8255,
clip marker paths=true,
ytick style={color=darkgray176}
]
\path [draw=mediumpurple148103189, draw opacity=0.7, semithick]
(axis cs:375.215330350027,512.4)
--(axis cs:484.984669649973,512.4);

\path [draw=mediumpurple148103189, draw opacity=0.7, semithick]
(axis cs:430.1,463.258132998341)
--(axis cs:430.1,561.541867001659);

\path [draw=mediumpurple148103189, draw opacity=0.7, semithick]
(axis cs:178.302230992387,114.8)
--(axis cs:235.497769007613,114.8);

\path [draw=mediumpurple148103189, draw opacity=0.7, semithick]
(axis cs:206.9,102.493672009393)
--(axis cs:206.9,127.106327990607);

\path [draw=darkorange2301300, draw opacity=0.7, semithick]
(axis cs:2229.90576113399,23.15)
--(axis cs:2822.09423886601,23.15);

\path [draw=darkorange2301300, draw opacity=0.7, semithick]
(axis cs:2526,20.3398622201852)
--(axis cs:2526,25.9601377798148);

\path [draw=deepskyblue0176240, draw opacity=0.7, semithick]
(axis cs:1041.85054049562,79.86)
--(axis cs:1328.14945950438,79.86);

\path [draw=deepskyblue0176240, draw opacity=0.7, semithick]
(axis cs:1185,71.7156951726904)
--(axis cs:1185,88.0043048273096);

\path [draw=mediumpurple148103189, draw opacity=0.7, semithick]
(axis cs:205.838752276016,682.1)
--(axis cs:269.361247723984,682.1);

\path [draw=mediumpurple148103189, draw opacity=0.7, semithick]
(axis cs:237.6,616.804217834531)
--(axis cs:237.6,747.395782165469);

\path [draw=mediumpurple148103189, draw opacity=0.7, semithick]
(axis cs:354.51290045662,669.8)
--(axis cs:468.88709954338,669.8);

\path [draw=mediumpurple148103189, draw opacity=0.7, semithick]
(axis cs:411.7,603.704193193624)
--(axis cs:411.7,735.895806806376);

\path [draw=mediumpurple148103189, draw opacity=0.7, semithick]
(axis cs:316.524369077286,112.3)
--(axis cs:413.675630922714,112.3);

\path [draw=mediumpurple148103189, draw opacity=0.7, semithick]
(axis cs:365.1,100.05424334846)
--(axis cs:365.1,124.54575665154);

\path [draw=deepskyblue0176240, draw opacity=0.7, semithick]
;

\path [draw=deepskyblue0176240, draw opacity=0.7, semithick]
;

\path [draw=deepskyblue0176240, draw opacity=0.7, semithick]
;

\path [draw=deepskyblue0176240, draw opacity=0.7, semithick]
;

\path [draw=darkorange2301300, draw opacity=0.7, semithick]
(axis cs:1764.3215024548,24.97)
--(axis cs:2223.6784975452,24.97);

\path [draw=darkorange2301300, draw opacity=0.7, semithick]
(axis cs:1994,22.0258552531258)
--(axis cs:1994,27.9141447468742);

\path [draw=darkorange2301300, draw opacity=0.7, semithick]
(axis cs:3199.29482712257,31.25)
--(axis cs:4032.70517287743,31.25);

\path [draw=darkorange2301300, draw opacity=0.7, semithick]
(axis cs:3616,27.303196861499)
--(axis cs:3616,35.196803138501);

\addplot [semithick, mediumpurple148103189, opacity=0.7, mark=*, mark size=3, mark options={solid}, only marks, forget plot]
table {%
430.1 512.4
};
\addplot [semithick, mediumpurple148103189, opacity=0.7, mark=*, mark size=3, mark options={solid}, only marks, forget plot]
table {%
206.9 114.8
};
\addplot [semithick, darkorange2301300, opacity=0.7, mark=*, mark size=3, mark options={solid}, only marks, forget plot]
table {%
2526 23.15
};
\addplot [semithick, deepskyblue0176240, opacity=0.7, mark=*, mark size=3, mark options={solid}, only marks, forget plot]
table {%
1185 79.86
};
\addplot [semithick, mediumpurple148103189, opacity=0.7, mark=*, mark size=3, mark options={solid}, only marks, forget plot]
table {%
237.6 682.1
};
\addplot [semithick, mediumpurple148103189, opacity=0.7, mark=*, mark size=3, mark options={solid}, only marks, forget plot]
table {%
411.7 669.8
};
\addplot [semithick, mediumpurple148103189, opacity=0.7, mark=*, mark size=3, mark options={solid}, only marks, forget plot]
table {%
365.1 112.3
};
\addplot [semithick, deepskyblue0176240, opacity=0.7, mark=*, mark size=3, mark options={solid}, only marks, forget plot]
table {%
0 0
};
\addplot [semithick, deepskyblue0176240, opacity=0.7, mark=*, mark size=3, mark options={solid}, only marks, forget plot]
table {%
0 0
};
\addplot [semithick, darkorange2301300, opacity=0.7, mark=*, mark size=3, mark options={solid}, only marks, forget plot]
table {%
1994 24.97
};
\addplot [semithick, darkorange2301300, opacity=0.7, mark=*, mark size=3, mark options={solid}, only marks, forget plot]
table {%
3616 31.25
};

\addlegendimage{deepskyblue0176240, mark=* , only marks, mark size=3}
\addlegendimage{darkorange2301300, mark= *,  only marks, mark size=3}
\addlegendimage{mediumpurple148103189, mark= *, only marks, mark size=3}

\legend{SS,SS+Li,C+W}
\end{axis}
\end{tikzpicture}

%% file: DvsLideposited.tex
\begin{tikzpicture}

\definecolor{darkgray176}{RGB}{176,176,176}
\definecolor{darkorange2301300}{RGB}{230,130,0}
\definecolor{deepskyblue0176240}{RGB}{0,176,240}
\definecolor{lightgray204}{RGB}{204,204,204}
\definecolor{mediumpurple148103189}{RGB}{148,103,189}

\begin{axis}[
legend columns=1,
legend style={
    /tikz/column 2/.style={
        column sep=3pt,
    },
  fill opacity=0.8,
  draw opacity=1,
  text opacity=1,
  at={(0.3,0.91)},
  clip marker paths=true,
  anchor=north,
  draw=lightgray204
},tick align=inside,
tick pos=left,
x grid style={darkgray176},
xlabel={In situ deposited Li areal density \(\displaystyle [\mathrm{10^{15}at/cm^{2}]}\)},
xmin=0, xmax=1306.48,
xtick style={color=darkgray176},
y grid style={darkgray176},
ylabel={D areal density \(\displaystyle [\mathrm{10^{15}at/cm^{2}]}\)},
ymin=0, ymax=87.8255,
ytick style={color=darkgray176}
]

\path [draw=darkorange2301300, draw opacity=0.7, semithick]
(axis cs:777.122957477705,23.15)
--(axis cs:946.877042522295,23.15);

\path [draw=darkorange2301300, draw opacity=0.7, semithick]
(axis cs:862,20.3398622201852)
--(axis cs:862,25.9601377798148);

\path [draw=deepskyblue0176240, draw opacity=0.7, semithick]
(axis cs:1041.85054049562,79.86)
--(axis cs:1328.14945950438,79.86);

\path [draw=deepskyblue0176240, draw opacity=0.7, semithick]
(axis cs:1185,71.7156951726904)
--(axis cs:1185,88.0043048273096);

\path [draw=deepskyblue0176240, draw opacity=0.7, semithick]
(axis cs:0,0)
--(axis cs:0,0);

\path [draw=deepskyblue0176240, draw opacity=0.7, semithick]
;

\path [draw=deepskyblue0176240, draw opacity=0.7, semithick]
(axis cs:0,0)
--(axis cs:0,0);

\path [draw=deepskyblue0176240, draw opacity=0.7, semithick]
;

\path [draw=darkorange2301300, draw opacity=0.7, semithick]
(axis cs:303.156665282521,24.97)
--(axis cs:356.843334717479,24.97);

\path [draw=darkorange2301300, draw opacity=0.7, semithick]
(axis cs:330,22.0258552531258)
--(axis cs:330,27.9141447468742);

\path [draw=darkorange2301300, draw opacity=0.7, semithick]
(axis cs:150.753129666776,31.25)
--(axis cs:217.246870333224,31.25);

\path [draw=darkorange2301300, draw opacity=0.7, semithick]
(axis cs:184,27.303196861499)
--(axis cs:184,35.196803138501);

\addplot [thick, dashed, gray, opacity=0.3,forget plot]
table {%
0 4.92222108580341
0 4.92222108580341
184 14.2405295283153
330 21.6344047055259
862 48.5764704197451
1185 64.9341531748069
};
\addplot [semithick, darkorange2301300, opacity=0.7, mark=*, mark size=3, mark options={solid}, only marks,forget plot]
table {%
862 23.15
};
\addplot [semithick, deepskyblue0176240, opacity=0.7, mark=*, mark size=3, mark options={solid}, only marks,forget plot]
table {%
1185 79.86
};
\addplot [semithick, deepskyblue0176240, opacity=0.7, mark=*, mark size=3, mark options={solid}, only marks,forget plot]
table {%
0 0
};
\addplot [semithick, deepskyblue0176240, opacity=0.7, mark=*, mark size=3, mark options={solid}, only marks,forget plot]
table {%
0 0
};
\addplot [semithick, darkorange2301300, opacity=0.7, mark=*, mark size=3, mark options={solid}, only marks,forget plot]
table {%
330 24.97
};
\addplot [semithick, darkorange2301300, opacity=0.7, mark=*, mark size=3, mark options={solid}, only marks,forget plot]
table {%
184 31.25
};

\addlegendimage{deepskyblue0176240, mark=* , only marks, mark size=3}
\addlegendimage{darkorange2301300, mark= *,  only marks, mark size=3}

\legend{SS,SS+Li}
\end{axis}
\end{tikzpicture}

%% file: main.bbl
\begin{thebibliography}{10}

\bibitem{you2022divertor}
JH~You, G~Mazzone, E~Visca, H~Greuner, M~Fursdon, Y~Addab, C~Bachmann, T~Barrett, U~Bonavolont{\`a}, B~B{\"o}swirth, et~al.
\newblock Divertor of the european demo: Engineering and technologies for power exhaust.
\newblock {\em Fusion Engineering and Design}, 175:113010, 2022.

\bibitem{romanelli2012fusion}
Francesco Romanelli, P~Barabaschi, D~Borba, G~Federici, L~Horton, R~Neu, D~Stork, and H~Zohm.
\newblock Fusion electricity: A roadmap to the realization of fusion energy.
\newblock 2012.

\bibitem{rodriguez2022overview}
P~Rodriguez-Fernandez, AJ~Creely, MJ~Greenwald, D~Brunner, SB~Ballinger, CP~Chrobak, DT~Garnier, R~Granetz, ZS~Hartwig, NT~Howard, et~al.
\newblock Overview of the sparc physics basis towards the exploration of burning-plasma regimes in high-field, compact tokamaks.
\newblock {\em Nuclear Fusion}, 62(4):042003, 2022.

\bibitem{loarte2024initial}
A~Loarte, RA~Pitts, T~Wauters, I~Nunes, F~K{\"o}chl, AR~Polevoi, SH~Kim, M~Lehnen, M~Schneider, L~Zabeo, et~al.
\newblock Initial evaluations in support of the new iter baseline and research plan.
\newblock 2024.

\bibitem{zamperini2023separatrix}
Shawn Zamperini, Tyler Abrams, JH~Nichols, JD~Elder, Jonah~D Duran, PC~Stangeby, DC~Donovan, DL~Rudakov, Andreas Wingen, and C~Crowe.
\newblock Separatrix-to-wall simulations of impurity transport with a fully three-dimensional wall in diii-d.
\newblock {\em Fusion Science and Technology}, 79(1):36--45, 2023.

\bibitem{nygren2016liquid}
RE~Nygren and FL~Tabar{\'e}s.
\newblock Liquid surfaces for fusion plasma facing components—a critical review. part i: Physics and psi.
\newblock {\em Nuclear Materials and Energy}, 9:6--21, 2016.

\bibitem{tabares2015present}
Francisco~L Tabar{\'e}s.
\newblock Present status of liquid metal research for a fusion reactor.
\newblock {\em Plasma Physics and Controlled Fusion}, 58(1):014014, 2015.

\bibitem{morgan2017liquid}
Thomas~W Morgan, Peter Rindt, GG~Van~Eden, Vladimir Kvon, MA~Jaworksi, and NJ~Lopes Cardozo.
\newblock Liquid metals as a divertor plasma-facing material explored using the pilot-psi and magnum-psi linear devices.
\newblock {\em Plasma Physics and Controlled Fusion}, 60(1):014025, 2017.

\bibitem{ono2017liquid}
M~Ono, MA~Jaworski, R~Kaita, Y~Hirooka, TK~Gray, NSTX-U~Research Team, et~al.
\newblock Liquid lithium applications for solving challenging fusion reactor issues and nstx-u contributions.
\newblock {\em Fusion Engineering and Design}, 117:124--129, 2017.

\bibitem{maingi2011continuous}
Rajesh Maingi, SM~Kaye, CH~Skinner, DP~Boyle, JM~Canik, MG~Bell, RE~Bell, Travis~K Gray, MA~Jaworski, R~Kaita, et~al.
\newblock Continuous improvement of h-mode discharge performance with progressively increasing lithium coatings in the national spherical torus experiment.
\newblock {\em Physical review letters}, 107(14):145004, 2011.

\bibitem{sun2019real}
Z~Sun, R~Maingi, JS~Hu, W~Xu, GZ~Zuo, YW~Yu, CR~Wu, M~Huang, XC~Meng, L~Zhang, et~al.
\newblock Real time wall conditioning with lithium powder injection in long pulse h-mode plasmas in east with tungsten divertor.
\newblock {\em Nuclear Materials and Energy}, 19:124--130, 2019.

\bibitem{osborne2015enhanced}
Thomas~H Osborne, Gary~L Jackson, Zheng Yan, Rajesh Maingi, Dennis~K Mansfield, Brian~A Grierson, Chris~P Chrobak, Adam~G McLean, Steve~L Allen, Devon~J Battaglia, et~al.
\newblock Enhanced h-mode pedestals with lithium injection in diii-d.
\newblock {\em Nuclear Fusion}, 55(6):063018, 2015.

\bibitem{effenberg2022mitigation}
Florian Effenberg, Alessandro Bortolon, Livia Casali, Raffi Nazikian, Igor Bykov, Filippo Scotti, Huiqian~Q Wang, Max~E Fenstermacher, Robert Lunsford, Alexander Nagy, et~al.
\newblock Mitigation of plasma--wall interactions with low-z powders in diii-d high confinement plasmas.
\newblock {\em Nuclear Fusion}, 62(10):106015, 2022.

\bibitem{boyle2023extending}
Dennis~P Boyle, J~Anderson, S~Banerjee, RE~Bell, W~Capecchi, DB~Elliott, C~Hansen, S~Kubota, BP~LeBlanc, A~Maan, et~al.
\newblock Extending the low-recycling, flat temperature profile regime in the lithium tokamak experiment-$\beta$ (ltx-$\beta$) with ohmic and neutral beam heating.
\newblock {\em Nuclear Fusion}, 63(5):056020, 2023.

\bibitem{baldwin2002deuterium}
MJ~Baldwin, RP~Doerner, SC~Luckhardt, and RW~Conn.
\newblock Deuterium retention in liquid lithium.
\newblock {\em Nuclear fusion}, 42(11):1318, 2002.

\bibitem{veleckis1974lithium}
Ewald Veleckis, Erven~H Van~Deventer, and Milton Blander.
\newblock Lithium-lithium hydride system.
\newblock {\em The Journal of Physical Chemistry}, 78(19):1933--1940, 1974.

\bibitem{morbey2024deuterium}
M~Morbey, J~Gonzalez, WM~Arnoldbik, B~Tyburska-Pueschel, and TW~Morgan.
\newblock Deuterium retention in co-deposition with lithium in magnum-psi: experimental analysis and comparison with solps-iter.
\newblock {\em Nuclear Fusion}, 64(7):076019, 2024.

\bibitem{ou2022deuterium}
W~Ou, WM~Arnoldbik, K~Li, P~Rindt, and TW~Morgan.
\newblock Deuterium retention and removal in liquid lithium determined by in situ nra in magnum-psi.
\newblock {\em Nuclear Fusion}, 62(7):076010, 2022.

\bibitem{paneta2012differential}
V~Paneta, A~Kafkarkou, M~Kokkoris, and A~Lagoyannis.
\newblock Differential cross-section measurements for the 7li (p, p0) 7li, 7li (p, p1) 7li, 7li (p, $\alpha$0) 4he, 19f (p, p0) 19f, 19f (p, $\alpha$0) 16o and 19f (p, $\alpha$1, 2) 16o reactions.
\newblock {\em Nuclear Instruments and Methods in Physics Research Section B: Beam Interactions with Materials and Atoms}, 288:53--59, 2012.

\bibitem{gurbich2016sigmacalc}
AF~Gurbich.
\newblock Sigmacalc recent development and present status of the evaluated cross-sections for iba.
\newblock {\em Nuclear Instruments and Methods in Physics Research Section B: Beam Interactions with Materials and Atoms}, 371:27--32, 2016.

\bibitem{wielunska2016cross}
B~Wielunska, M~Mayer, T~Schwarz-Selinger, U~Von~Toussaint, and J~Bauer.
\newblock Cross section data for the d (3he, p) 4he nuclear reaction from 0.25 to 6 mev.
\newblock {\em Nuclear Instruments and Methods in Physics Research Section B: Beam Interactions with Materials and Atoms}, 371:41--45, 2016.

\bibitem{bondouk1975experimental}
II~Bondouk, F~Asfour, Z~Saleh, and F~Machali.
\newblock An experimental investigation of the reactions 7 li (3 he, p 0) 9 be and 7 li (3 he, p 2) 9 be in the 3 he energy range of 1.0 to 2.5 mev.
\newblock {\em Revue Roumaine de Physique}, 20(10):1095--1098, 1975.

\bibitem{barradas1997simulated}
NP~Barradas, C~Jeynes, and RP~Webb.
\newblock Simulated annealing analysis of rutherford backscattering data.
\newblock {\em Applied Physics Letters}, 71(2):291--293, 1997.

\bibitem{matvejivcek2013influence}
Ji{\v{r}}{\'\i} Mat{\v{e}}j{\'\i}{\v{c}}ek, Monika Vil{\'e}mov{\'a}, Radek Mu{\v{s}}{\'a}lek, Pavel Sachr, and Jakub Horn{\'\i}k.
\newblock The influence of interface characteristics on the adhesion/cohesion of plasma sprayed tungsten coatings.
\newblock {\em Coatings}, 3(2):108--125, 2013.

\bibitem{van2020effect}
JPB Van~Dam, ST~Abrahami, A~Yilmaz, Y~Gonzalez-Garcia, H~Terryn, and JMC Mol.
\newblock Effect of surface roughness and chemistry on the adhesion and durability of a steel-epoxy adhesive interface.
\newblock {\em International Journal of Adhesion and Adhesives}, 96:102450, 2020.

\bibitem{lao1985reconstruction}
LL~Lao, H~St John, RD~Stambaugh, AG~Kellman, and W~Pfeiffer.
\newblock Reconstruction of current profile parameters and plasma shapes in tokamaks.
\newblock {\em Nuclear fusion}, 25(11):1611, 1985.

\bibitem{nagy2018multi}
A~Nagy, A~Bortolon, DM~Mauzey, E~Wolfe, EP~Gilson, R~Lunsford, R~Maingi, DK~Mansfield, R~Nazikian, and AL~Roquemore.
\newblock A multi-species powder dropper for magnetic fusion applications.
\newblock {\em Review of Scientific Instruments}, 89(10), 2018.

\bibitem{MANSFIELD2010890}
D.K. Mansfield, A.L. Roquemore, H.~Schneider, J.~Timberlake, H.~Kugel, and M.G. Bell.
\newblock A simple apparatus for the injection of lithium aerosol into the scrape-off layer of fusion research devices.
\newblock {\em Fusion Engineering and Design}, 85(6):890--895, 2010.
\newblock Proceedings of the 1st International Workshop on Lithium Applications for the Boundary Control in Fusion Devices.

\bibitem{bortolon2020observations}
A~Bortolon, R~Maingi, A~Nagy, J~Ren, JD~Duran, A~Maan, DC~Donovan, JA~Boedo, DL~Rudakov, AW~Hyatt, et~al.
\newblock Observations of wall conditioning by means of boron powder injection in diii-d h-mode plasmas.
\newblock {\em Nuclear Fusion}, 60(12):126010, 2020.

\bibitem{EFFENBERG2025101832}
F.~Effenberg, K.~Schmid, F.~Nespoli, A.~Bortolon, Y.~Feng, B.A. Grierson, J.D. Lore, R.~Maingi, and D.L. Rudakov.
\newblock Integrated modeling of boron powder injection for real-time plasma-facing component conditioning.
\newblock {\em Nuclear Materials and Energy}, 42:101832, 2025.

\bibitem{wong1992divertor}
CPC Wong, R~Junge, RD~Phelps, P~Politzer, F~Puhn, WP~West, R~Bastasz, D~Buchenauer, W~Hsu, J~Brooks, et~al.
\newblock Divertor materials evaluation system at diii-d.
\newblock {\em Journal of nuclear materials}, 196:871--875, 1992.

\bibitem{watkins2008high}
JG~Watkins, D~Taussig, RL~Boivin, MA~Mahdavi, and RE~Nygren.
\newblock High heat flux langmuir probe array for the diii-d divertor plates.
\newblock {\em Review of scientific instruments}, 79(10), 2008.

\bibitem{brooks1992multichord}
NH~Brooks, A~Howald, K~Klepper, and P~West.
\newblock Multichord spectroscopy of the diii-d divertor region.
\newblock {\em Review of scientific instruments}, 63(10):5167--5169, 1992.

\bibitem{fonck1982multichannel}
RJ~Fonck, AT~Ramsey, and RV~Yelle.
\newblock Multichannel grazing-incidence spectrometer for plasma impurity diagnosis: Spred.
\newblock {\em Applied Optics}, 21(12):2115--2123, 1982.

\bibitem{mclean2018quantification}
AG~McLean, SL~Allen, ME~Fenstermacher, AE~Jaervinen, CJ~Lasnier, GD~Porter, T~Rognlien, CM~Samuell, VA~Soukhanovskii, J~Boedo, et~al.
\newblock Quantification of radiating species in the diii-d divertor in the transition to detachment using extreme ultraviolet spectroscopy.
\newblock In {\em Proceedings of the 27th IAEA Fusion Energy Conference in Gandhinagar, India}, 2018.

\bibitem{pospieszczyk2010determination}
A~Pospieszczyk, D~Borodin, S~Brezinsek, A~Huber, A~Kirschner, Ph~Mertens, G~Sergienko, B~Schweer, IL~Beigman, and L~Vainshtein.
\newblock Determination of rate coefficients for fusion-relevant atoms and molecules by modelling and measurement in the boundary layer of textor.
\newblock {\em Journal of Physics B: Atomic, Molecular and Optical Physics}, 43(14):144017, 2010.

\bibitem{abrams2021evaluation}
Tyler Abrams, S~Bringuier, Dan~M Thomas, Gregory Sinclair, S~Gonderman, L~Holland, Dmitry~L Rudakov, Robert~S Wilcox, Ezekial~A Unterberg, and Filippo Scotti.
\newblock Evaluation of silicon carbide as a divertor armor material in diii-d h-mode discharges.
\newblock {\em Nuclear Fusion}, 61(6):066005, 2021.

\bibitem{summers2007adas}
HP~Summers, MG~O’Mullane, AD~Whiteford, NR~Badnell, and SD~Loch.
\newblock Adas: Atomic data, modelling and analysis for fusion.
\newblock In {\em Aip conference proceedings}, volume 901, pages 239--248. American Institute of Physics, 2007.

\bibitem{naujoks1996tungsten}
D~Naujoks, K~Asmussen, M~Bessenrodt-Weberpals, S~Deschka, R~Dux, W~Engelhardt, AR~Field, G~Fussmann, JC~Fuchs, C~Garcia-Rosales, et~al.
\newblock Tungsten as target material in fusion devices.
\newblock {\em Nuclear fusion}, 36(6):671, 1996.

\bibitem{stangeby2000plasma}
Peter~C Stangeby et~al.
\newblock {\em The plasma boundary of magnetic fusion devices}, volume 224.
\newblock Institute of Physics Pub. Philadelphia, Pennsylvania, 2000.

\bibitem{behrisch2007sputtering}
Rainer Behrisch and Wolfgang Eckstein.
\newblock {\em Sputtering by particle bombardment: experiments and computer calculations from threshold to MeV energies}, volume 110.
\newblock Springer Science \& Business Media, 2007.

\bibitem{abrams2014erosion}
T~Abrams, MA~Jaworski, R~Kaita, DP~Stotler, G~De~Temmerman, TW~Morgan, MA~Van Den~Berg, and HJ~Van Der~Meiden.
\newblock Erosion of lithium coatings on tzm molybdenum and graphite during high-flux plasma bombardment.
\newblock {\em Fusion Engineering and Design}, 89(12):2857--2863, 2014.

\bibitem{effenberg2023situ}
Florian Effenberg, Shota Abe, Gregory Sinclair, Tyler Abrams, Alessandro Bortolon, William~R Wampler, Florian~M Laggner, Dmitry~L Rudakov, Igor Bykov, Charles~J Lasnier, et~al.
\newblock In-situ coating of silicon-rich films on tokamak plasma-facing components with real-time si material injection.
\newblock {\em Nuclear Fusion}, 63(10):106004, 2023.

\bibitem{evans2006edge}
Todd~E Evans, Richard~A Moyer, Keith~H Burrell, Max~E Fenstermacher, Ilon Joseph, Anthony~W Leonard, Thomas~H Osborne, Gary~D Porter, Michael~J Schaffer, Philip~B Snyder, et~al.
\newblock Edge stability and transport control with resonant magnetic perturbations in collisionless tokamak plasmas.
\newblock {\em nature physics}, 2(6):419--423, 2006.

\bibitem{hager2019gyrokinetic}
Robert Hager, CS~Chang, NM~Ferraro, and R~Nazikian.
\newblock Gyrokinetic study of collisional resonant magnetic perturbation (rmp)-driven plasma density and heat transport in tokamak edge plasma using a magnetohydrodynamic screened rmp field.
\newblock {\em Nuclear fusion}, 59(12):126009, 2019.

\bibitem{goldston2016lithium}
Robert~James Goldston, Rachel Myers, and Jacob Schwartz.
\newblock The lithium vapor box divertor.
\newblock {\em Physica Scripta}, 2016(T167):014017, 2016.

\end{thebibliography}
